# On classification approaches for crystallographic symmetries of noisy 2D periodic patterns

Peter Moeck

*Abstract* — **The existing types of classification approaches for the crystallographic symmetries of patterns that are more or less periodic in two dimensions (2D) are reviewed. Their relative performance is evaluated in a qualitative manner. Pseudo-symmetries of different kinds are discussed as they present severe challenges to most classification approaches when noise levels are moderate to high. The author's information theory based approaches utilize digital images and geometric Akaike Information Criteria. They perform well in the presence of pseudo-symmetries and turn out to be the only ones that allow for fully objective (completely researcher independent) and generalized noise level dependent classifications of the full range of crystallographic symmetries, i.e. Bravais lattice type, Laue class, and plane symmetry group, of noisy real-world images. His method's identification of the plane symmetry group that can with the highest likelihood be assigned to a noisy 2D periodic image enables the most meaningful crystallographic averaging in the spatial frequency domain. This kind of averaging suppresses generalized noise much more effectively than traditional Fourier filtering. Taking account of the fact that it is fundamentally unsound to assign an abstract mathematical concept such as a single symmetry type, class, or group with 100 % certainty to a more or less 2D periodic record of a noisy real-world imaging experiment that involved a real-world sample, the author's information theory based approaches to crystallographic symmetry classifications deliver probabilistic (rather than definitive) classifications. Recent applications of deep convolutional neural networks (DCNNs) to classifications of crystallographic translation symmetry types in 2D and crystals in three dimensions (3D) are discussed as these "correlation detection and optimization" machines deliver probabilistic classifications by other – non-analytical – means. The discussed DCNN classifications ignore the fact that many crystallographic symmetries are hierarchic, i.e. that the classification classes are often non-disjoint. They are currently also incapable of dealing with pseudo-symmetries. DCNNs for classifications of crystals in 3D are discussed separately in an appendix.**

## I. Introduction

The mathematical framework for the classification of crystallographic symmetries [1-4] that are compatible with translation periodicity in two dimensions (2D) was derived independently by a Russian polymath [5] and a Hungarian mathematician [6] (while the latter was consulting with the eminent crystallographer and geoscientist Paul Niggli in Switzerland) in the years 1891 and 1924, respectively.

This work was supported by a Faculty Enhancement Grant of Portland State University. Peter Moeck is with the Nano-Crystallography Group of the Department of Physics, Portland State University, Portland, Oregon, USA, (e-mail: pmoeck@pdx.edu).

The derived mathematical framework is nowadays utilized for the classification of (*i*) highly magnified but necessarily noisy experimental images from crystals that are sufficiently thin (so that quasi-kinematic imaging conditions prevail) and crystal surfaces as recorded with modern transmission electron and scanning probe microscopes [7-18], of (*ii*) both perfectly 2D periodic and noisy synthetic images [18-26], as well as of (*iii*) more or less 2D periodic images from rugs, windows, a honeycomb, both a tiled floor and wall, metal gates, a metallic anti-skid surface [26], industrially manufactured wallpapers [27-29], textiles, and decorative ceramic tiles [29-33], medieval Islamic building ornaments [34], and images of (*iv*) many other objects and scenes [35].

Strictly speaking, crystallographic symmetry classifications in 2D require an infinitely large image with an infinite number of unit cell repeats, but it is customary to ignore both the finite nature of more or less 2D periodic images and the finite number of their repeat units for the sake of the practicality of the symmetry classification. Also for the sake of practicality, one typically accepts the consequences of the facts that the pixel size in a digital image is not infinitely small and that the total number of pixels is not infinitely large.

Some of the above-mentioned noise-free synthetic images and noisy images of everyday objects [26], wallpapers [27-29], textiles and ceramic tiles [29-33], as well as those from Islamic building ornaments [34] were classified in direct/physical space with respect to their Bravais lattice type and plane symmetry group, as it is customary in both the computational symmetry community [35] and the community that studies Islamic building ornaments either visually in the field [36,37] or on their computers [34,35].

All of the other above-mentioned images were classified in reciprocal/Fourier space with respect to their Bravais lattice type and plane symmetry group as it is customary in 2D crystallography, transmission electron microscopy, and scanning probe microscopy communities [7-23]. The above-mentioned framework for the classification of crystallographic symmetries [1-5] in 2D is obviously applicable regardless of the space in which the classifications are made.

Similarly obvious is the fact that members of the communities that traditionally perform direct space classifications may benefit from becoming informed about the opportunities that the Fourier/reciprocal space offers with respect to crystallographic symmetry classifications of more or less 2D periodic images [7-23]. Members of the 2D crystallography, transmission electron microscopy, and scanning probe microscopy communities are conversely bound to benefit from an analysis of what has been achieved so far in direct space classification of the crystallographic symmetries of more or less 2D periodic images [26-35]. Approaching the same kinds of problems in an analytical manner and in two complementary spaces while using the same mathematical framework [1-5] calls for the exchange of ideas between scientists from different communities. The present paper aims at fostering such exchanges, which may eventually lead to collaborations for the greater good.

The immediately following section of this paper discusses several of the analytical classification approaches for the crystallographic symmetries of more or less 2D periodic images in both direct and reciprocal space. Providing such a review is one of the goals of this paper. The discussions in the second section of this paper will reveal the non-obvious "nature" of crystallographic symmetry classification problems and identify the pitfalls that need to be addressed by all of the individual approaches.

The author's recently developed information theory based approaches to crystallographic symmetry classifications [16-23] utilize geometric bias-corrected sums of squared residuals in the form of geometric Akaike Information Criteria [38-42]. His approach to plane symmetry group classifications [16,17,21-23] will be discussed in the second section of this paper in some more detail as part of the review of the analytical approaches. The discussions will proceed at a qualitative level as all of the other approaches are also dealt with in an essentially qualitative manner.

The author's other crystallographic symmetry classification approaches, i.e. into Bravais lattice types [18-20] and crystallographic [16,23] as well as non-crystallographic Laue classes, are discussed in some detail elsewhere as they serve other purposes than identifying the plane symmetry group that can with the highest likelihood be assigned to a noisy 2D periodic image. Note that it is this particular identification that enables the most meaningful crystallographic averaging in the spatial frequency domain that is of paramount interest to 2D crystallographers, transmission electron and scanning probe microscopists.

There will, however, be some generalizations in the second section of this paper when general aspects of the author's information theory based approaches to crystallographic symmetry classifications are concerned. Possible alternatives to the author's approaches could employ different geometric information criteria and will be briefly discussed in the penultimate section of this paper.

There is finally one further distinction to be made with respect to the author's information theory based approaches. The plane symmetries (and any other crystallographic or non-crystallographic symmetries) into which a noisy 2D periodic (or quasiperiodic) image is to be classified can either be within the same symmetry hierarchy[1] branch [16-18], i.e. non-disjoint, or within different symmetry hierarch branches [21,22], i.e. disjoint.

Most of the experienced practitioners of 2D electron crystallography, e.g. [13,14], and microscopists, e.g. [11,12], have so far not embraced information theory based approaches to crystallographic symmetry classifications in their work but could do so to their benefit. Most of the computer scientists that work on "computational symmetry" studies [35] have so far also not embraced these approaches, but could likewise do so as there are no known drawbacks. The latter scientists may like to take up a deceptively simple idea from the first paragraph of the second section of this paper that would allow them to work in a computationally efficient manner in Fourier/reciprocal space. To alert the members of both types of communities to the opportunities of information theory based approaches to crystallographic symmetry classifications in Fourier space is the main reason for publishing this review at the present time.

The two other goals of this paper are to (*i*) illustrate the nature of severe pseudo-symmetry challenges that the author's information theory based approach to plane symmetry group classifications [17,21,22] overcome [23], and to (*ii*) contrast the *analytical* approaches [7-45] with the "brute calculation power" *probabilistic* approaches of employing deep convolutional neural networks (DCNNs) to selected aspects of crystallographic symmetry [46-48] and crystalline compound [49] classification tasks.

Pseudo-symmetries of the Fedorov type [43] as well as metric and motif based pseudo-symmetries [44] will be discussed in the third and fourth sections of this paper. The basis of these discussions will be two pairs of synthetic images, whereby (*i*) one image of each pair is free of noise and (*ii*) Gaussian distributed noise of mean zero has been added to that image in order to create a more or less 2D periodic image for demonstration purposes. Such discussions within this paper are deemed necessary because the real-world existence of pseudo-symmetries has already led to many questionable and wrong entries in major crystallographic databases, as discussed in some detail in [22].

In order to achieve the second of the above mentioned goals, the fifth section of this paper provides a brief discussion of the state of the art of the application of such a "non-linear regression based correlation detection machine" to translation symmetry classifications in 2D [46]. Since experimental scanning tunneling microscope images from that study were put into open access (and are freely available in the on-line supporting material of [46]), this author reproduced one of the more interesting classification results of the DCNN in [46] with his information theory based approach to plane symmetry group classifications [16]. The paper ends with a summary and conclusions section.

An appendix discusses the state of the art in classifications of selected aspects of crystal symmetries in 3D by means of DCNNs [47,48] and provides the basis for comparisons of their performance to that of the 2D translation symmetry classifier of [46] (as discussed in the fifth section of this paper). A classification of crystalline compounds by such machines that is aided by the morphologies of individual crystals [49] is also mentioned in the appendix.

The nature of the *probabilistic* crystallographic symmetry classifications by DCNNs [46-48] is very different from that of the *analytical* classifications (which have been used for over a century) so that they need to be discussed separately in conjunction with their more or less specific background [50-58]. The fifth section on this paper and the appendix serve two purposes.

[1] The crystallographic standard term for symmetry hierarchy branch is "chain of maximal subgroups", see [1] for mathematical definitions. A crucial distinction is to be made between translationengleiche and klassengleiche subgroups. The former subgroups are characterized by the retention of all translations while one descents from a minimal supergroup to a maximal subgroup. In other words, the area of a 2D unit cell remains the same when one "moves down" within the same symmetry hierarchy branch.

The first of these purposes is to alert the proponents of crystallographic symmetry classifications by means of DCNNs to the real-world existence of different types of pseudo-symmetries, the hierarchic nature of crystallographic symmetries, and the relevance of Kenichi Kanatani's deep statistics insights [38-42] to their endeavors (as discussed throughout sections II to IV). The second purpose is to correct the publication record as far as the usage of non-standard crystallographic terms in [47,48] is concerned.

Whereas the described achievements of DCNNs are quite impressive, fundamental problems concerning pseudo-symmetries and the hierarchic nature of crystallographic symmetries are so far ignored by the proponents of the DCNN approaches. The classifications that these machines produce are probabilistic, but can on a fundamental level only be "correct" when the labels on *all* of the training data have been assigned in a fully objective manner so that they are both noise level dependent and correct in the information theoretic sense at any one time. For reasons stated below, this is currently not the case.

Note in passing that the author's information theory based approaches to crystallographic symmetry classifications [16-23] are analytic in nature, but deliver probabilistic results with necessity.

## II. ANALYTICAL PLANE SYMMETRY GROUP CLASSIFICATIONS IN 2D

The images and object patterns that the members of the computational symmetry and Islamic building ornaments appreciation communities are concerned with contain typically only one or a few periodic repeats so that alternative Fourier/reciprocal space approaches to crystallographic symmetry classifications seem to be, at first sight, without merit. One can, however, "combine" the available periodic repeat(s) of these images into a single much larger image that features a sufficient number of repeats so that it can be classified successfully with respect to its crystallographic symmetries in Fourier space [59].

No new information is created by the stitching together of the unit cells but computationally efficient approaches to crystallographic symmetry classifications are enabled. In addition, *fully* objective, i.e. *completely* researcher independent approaches the crystallographic symmetry classifications become available when geometric Akaike Information Criteria are utilized (as in the author's more recent work on this subject [16-23]).

We used the freely available Microsoft program "Image Composite Editor" [60] for the creation of seamless combinations of unit cells of the three Islamic building ornament images that are in [34] presented as examples for which the original direct space classification failed to give results. These failures were due to (*i*) a low level of contrast in one of the images, (*ii*) two highly sophisticated motifs with interlacing[2], and (*iii*) comparatively large amounts of structural irregularities in the three building ornaments themselves. Our stitched-together versions of these three images presented no problem to their plane symmetry group classifications by means of the author's reciprocal/Fourier space approach [59].

The first of these three images in [34] features a comparatively high contrast, a motif with more or less straight edges, a rather high level of structural degradation, no interlacing, and approximately 16 repeats of the unit cell. Even without increasing the number of repeats in this image[3] with the above-mentioned stitching program, it could be classified in reciprocal/Fourier space with the author's information theory based approach [59].

A little more than four full unit cell repeats sufficed [59] for the successful application of the author's plane symmetry classification procedure when the translation periodic motif was of (*i*) high symmetry, the contrast in the image was (*ii*) large due to a rather uniform white background, and there were (*iii*) no major structural irregularities as in the image from a historic ceramic tile[4] in figure 2 of [31].

Systematic studies in high-resolution phase contrast transmission electron microscopy (HRTEM) based electron crystallography showed that one can obtain meaningful experimental results by means of crystallographic image processing [13] from processed image areas that contain as little as ten unit cells [14]. The (projected) plane symmetry group of a $Li_2NaTa_7O_{19}$ crystal has, for example, been derived correctly from a HRTEM image region that contained approximately 15 unit cells within a circular disk region with a diameter of 8.4 nm [14].

More representative electron crystallography results were, however, obtained from a circular disk area of the same HRTEM image when a four-fold larger region of that image, containing approximately 60 unit cells, was processed crystallographically. The Fourier coefficients of the image intensity within that four-fold larger image region were much better resolved as discrete peaks in the amplitude map of the calculated discrete Fourier transform [14]. The monograph on image-based electron crystallography [13] gives a few percent and less than 5° as typical error bars for the amplitudes and phase angles of the extracted Fourier coefficients when an inorganic crystal has been imaged at high resolution in a modern transmission electron microscope and an image region of about one hundred unit cells has been selected for the translation averaging that results from calculating the discrete Fourier transform of the image intensity. These kinds of error bars on the starting parameters are considered acceptable for the solving of a crystal structure, i.e. the finding of all atomic positions in the unit cell and asymmetric unit [13].

[2] A quasi-3D interlacing in the translation periodic motif is typically incompatible with the existence of genuine mirror-line symmetries. It leads to motif based pseudo-symmetries in 2D.

[3] That image was cut out of the downloadable row of three high-resolution images of figure 18 in the on-line version of [34]. It then contained 807 by 745 pixels squared. It was then padded into a 1024 by 1024 pixel image with an uniform background that consisted of the average gray scale level of the image. An elliptical region with half-axes of approximately 400 by 365 pixels was selected for the processing with the electron crystallography program CRISP [13,14]. Color information in the Islamic building ornament image was in the process of the classifications converted to 256 gray levels.

[4] That image is in color, of a size of 371 by 377 pixels, and freely downloadable from the researchgate.com version of [31]. It was padded into a 512 by 512 pixel image (by the same procedure as described in footnote 3) and a circular region with a diameter of approximately 370 pixels was finally selected for the processing with the electron crystallography program CRISP [13,14]. Color information was automatically converted to 256 gray levels by this program.

The author's own recommendation is that one should have some 50 to 100 unit cells in a noisy 2D periodic image (from a crystal or crystal surface under a microscope) in order to apply his information theory based classification methods for crystallographic symmetries. For images with high generalized noise levels, one is well advised to enter as many unit cells as possible into the analysis. Simply put, the more unit cells enter the procedure, i.e. the more information is there to be extracted and analyzed, the better the symmetry classification results will be.

Note that better results mean in this context higher confidence levels [17,18] of crystallographic symmetry classifications. For a synthetic noise-free image, the confidence level of a symmetry classification by one of the author's information theory based approaches would be very close to 100 %. The difference to 100 % would in such a case be due to the generalized noise quantity that is inadvertently introduced into the classification by actually performing [5] it with a computer program [39].

On a conceptual level, the seamless stitching together of one or a few unit cells to a much larger 2D periodic image with some 50 to 100 repeats would be similar to growing a crystal in order to derive the structure of the unit cell and asymmetric unit, whereby it is the presence of a large number of unit cell repeats that allows one to obtain the averaged crystal structure efficiently by means of Fourier analyses and syntheses [6] utilizing the translation symmetry restricted Euclidian 3D space group symmetries [1,2].

For more or less 2D periodic direct space images of everyday objects and scenes as well as Islamic building ornaments, the aims of crystallographic symmetry classifications are often to find the nearest plane symmetry group and Bravais lattice type. It needs to be noted here that the classifications often rely on the answers to series of dichotomous "yes or no" questions [24-26] if a certain symmetry operation is *visually* [36,37] or *computationally* [26-35] discernable in a real-world pattern or image.

As there is "generalized noise" in such images due to them having been recorded with some instrument under necessarily less than ideal conditions as well as approximations and finite numerical accuracy in all image processing algorithms [39,44] and because the objects and scenes belong to the real world, i.e. are actually never perfectly symmetric on a physical level, a dichotomous procedure cannot work in a logically strict[7] sense, but it is applied in the computational symmetry and Islamic building ornament appreciation communities anyway for the sake of its practicality when noise levels are low.

In cases of *visual* classifications of more or less 2D periodic patterns, e.g. Islamic building ornaments [36,37], there is no need for the recording of (noisy) images for the sake of a subsequent *computational* crystallographic symmetry classification. Generalized noise is, however, still a contingency of theses classifications since the patterns are not perfect [36] (in an abstract mathematical sense) and were degraded over time due to environmental influences and neglect by human beings. Arguably, these patterns were never perfect to begin with and structural defects are thought [36] to have on occasions been introduced intentionally.

There must, therefore, always be some degree of *subjectivity* involved in these kinds of crystallographic symmetry classifications. This subjectivity will obviously be some function of the generalized noise level, increase with that level, but decrease with both prior knowledge and the experience of the researcher that makes the classifications. Since personal value judgments that use some arbitrarily set thresholds will with necessity be involved, these kinds of symmetry classifications can never be fully objective.

The authors of [30-32,34], for example, had their computer program (named *"*FECETEX*"*) first classify all "object pieces" in the more or less 2D periodic repeat units of textiles, ceramic tile patterns, and medieval Islamic building ornaments with respect to their approximate point symmetries (higher than the identity rotation) and then combined the gathered information into an apparently well fitting plane symmetry group for the whole "ensemble" of object pieces that make up the whole image. The identified approximate point symmetries are, in effect, turned into perfectly obeyed crystallographic site symmetries [1,3]. It is almost needless to say that this cannot be done without employing *subjectively* set thresholds that are internal to the computer program. Some of these internal parameters are tabulated in [34]. Note that it is the relative closeness of the extracted real-world point symmetries in an image to the abstract mathematical site symmetries [1,3] that allows for the assignment of the abstract mathematical concepts of a plane symmetry group and Bravais lattice type to that image when the approach of [30-34] is employed.

The fact that certain crystallographic point symmetries are hierarchic, i.e. non-disjoint and are in supergroup-subgroup relationships, was not considered in [30-34]. Obviously, a certain site symmetry contains for the given geometric element all of the symmetry operations of its subgroups at the point symmetry level.

The resulting plane symmetry group classifications in the default setting of the FECETEX program will obviously be biased by the subjectivity that the computer programmers have put into the program's code. To make up for this, the computer programmers allow for a fine tuning of their parameters and threshold by the user as explicitly mentioned in [30,33]. Such fine tunings make the classifications, however, *not* completely *objective* either. A certain fine tuning may deliver excellent results with respect to the classification of a certain set of images only to fail in the classification of another set of images.

[5] Kenichi Kanatani stated in [39] that all computer programs will introduce some small quantity of generalized noise into any classification of any image. As a matter of fact, he developed geometric information criteria for the express purpose of dealing with both such inadvertent introductions of noise into computerized analyses and inclusion relations between classification classes.

[6] Note that there is no crystallographic phase problem associated with the analysis of more or less 2D periodic images from crystals in reciprocal space [13,14]. This is because there is no exclusive reliance on collapsed wave functions as experimentally obtainable from recorded diffraction patterns.

[7] In the strict logical sense, all local and global symmetries are broken by the presence of even the tiniest amount of uniformly distributed noise so that all "yes or no" questions would have to be answered by "no – not really".

For the apparently correct classification of 95 images from more or less 2D periodic images from textiles and ceramic tiles in [30], 64 % needed minor fine tunings of the program's internal parameters and 33 % needed major fine tunings. The remaining 3 % of these images were impossible to be classified into a reasonably well fitting plane symmetry group by the program in [30]. Unsurprisingly, these images where characterized by comparatively high generalized noise levels, high complexity at the motif level, and distributions of approximate point symmetries that could not be turned into site symmetries in a plane symmetry group in a consistent manner. Apparently without fine tuning, such inconsistencies prevented the program to make a classification into a plane symmetry group for 4 out of 19 more or less 2D periodic images from the Wikipedia entry for "Wallpaper groups" [34].

The computational costs of the individual stages of the plane symmetry group classification of an image with the FECETEX program are tabulated in [34]. Three of the listed 8 stages carry a computational cost on the order of the square of the number of objects that need to be classified with respect to their point/site symmetry groups. The computational cost of a plane symmetry group classification is for many more or less 2D periodic images much smaller when it is done in Fourier space. This is especially so when the image contains a large number of unit cells.

As already mentioned above, the author's plane symmetry classification method is based on the evaluation of ratios of geometric bias-corrected squared residuals between the raw and symmetrized Fourier coefficients of the image intensity. Quite independent of the number of pixels in a more or less 2D periodic image, typically a few tens to approximately a hundred Fourier coefficients contain all of the pertinent information on the translation averaged periodic motif. The computational cost of the author's information theory based plane symmetry group classification comprises only one stage that is on the order of the number of pixels times the logarithm (to the basis of 2) of that number, i.e. a single discrete Fourier transform. The so called "origin refinement" and the calculations of the ratios of the geometric bias-corrected squared residuals are all on the order of the number of the (complex-numbered) Fourier coefficients of the image intensity.

When there are many (more or less) straight edges in the motif, more Fourier coefficients will have an appreciable amplitude. In a typical high resolution image of a crystal or crystal surface from a microscope, there are no such edges so that there are fewer Fourier coefficients. Storing only the symmetrized Fourier coefficients of a more or less 2D periodic image, i.e. less than a hundred complex numbers, in a database is obviously much more effective than storing the whole symmetrized image at the pixel level.

In contrast to the dichotomous "yes or no" answer procedures outlined in [24-26], the computationally more demanding procedures of [27-34] allow for the identification of the origin of a crystallographic unit cell (of a plane symmetry group) as a byproduct. The origin identification capability is, however, not essential to these approaches since the standard crystallographic site symmetry information [1,3] that a correct origin choice would enable one to exploit is not explicitly used in [27-34].

Reference [27] states that noisy images with a large more or less uniform bright background and comparatively small dark repeating features are very difficult to classify in direct/physical space by any approach as the classification will depend on the intensity values of just the few pixels that comprise these features. The conspicuous elements that stand out from a more or less homogenous background in the approach of [30-34] and which are to be classified with respect to their 2D point symmetries as part of the plane symmetry group and Bravais lattice assignments to the image may also be composed of a small number of pixels. The smaller the number of pixels in these objects, the more ambiguous will be the objects' point/site symmetry classifications on the basis of some arbitrarily set thresholds.

Reciprocal space classifications of plane symmetry groups are, on the other hand, based on all pixels in a more or less 2D periodic image and deal with uniform backgrounds effectively by ignoring the amplitude of the 00 Fourier coefficient of the image intensity in their analyses of the prevailing plane symmetries so that they should be superior to the direct space approaches in this particular respect.

The summary and conclusions section of [34] ends with the idea that one could create a perfectly symmetric version of the analyzed image when all point/site symmetries higher than the identity and the prevailing translation symmetry type have been identified unambiguously. One could then enforce these symmetries on the totality of image pixels by averaging over the asymmetric units, but if there were mis-identifications due to the presence of generalized noise or the overlooked existence of pseudo-symmetries [43,44], this averaging would not lead to sensible results. The existing noise in the image would also be "symmetrized" along with the structure in the direct space of the image.

Better "symmetrizations" of noisy 2D periodic images have always been the desired outcome of the standard crystallographic image processing method that is used by the 2D crystallography, transmission electron microscopy, and scanning probe microscopy communities [7-23] and works in Fourier space. These symmetrizations combine the noise-removal feature of traditional Fourier filtering with the crystallographic averaging over all asymmetric units in the whole image, have been utilized for over 50 years, and contributed to the Nobel Prize in Chemistry to Sir Aaron Klug[8] in the year 1982.

The author's information theory based approach to plane symmetry group classifications [17,21,22] enables advancements on the state of the art in the crystallographic image processing field because it allows for the identification of the most meaningful separation of structure and generalized noise at the level of the asymmetric units. This kind of identification is, in turn, needed for the most meaningful symmetrization of a noisy 2D periodic image.

---

[8] Sir Aaron Klug's Nobel Prize citation reads: "*for his development of crystallographic electron microscopy and his structural elucidation of biologically important nucleic acid-protein complexes*".

Traditional Fourier filtering of such an image would only calculate the discrete Fourier transform of the image intensity, filter out all background intensity around the periodic-structure bearing Fourier coefficients that are laid out on the reciprocal lattice nodes in the amplitude map of this transform, and finally Fourier back-transform these coefficients into direct space. This is equivalent to the enforcement of the average translation symmetry on the image.

One can, therefore, equate traditional Fourier filtering with symmetrization in plane symmetry group *p1*. The crucial difference between traditional Fourier filtering and crystallographic image processing is the enforcement of plane symmetries higher than *p1* in the latter case. When the correct plane symmetry is enforced, one averages over the asymmetric unit, which can be up to 12 times smaller than the unit cell [1,3]. As there will be more entities to be averaged over, the averaging results will be more representative.

The vast majority of the generalized noise will not be correlated with the Fourier coefficients that represent the periodic structure in the image and will, therefore, effectively be filtered out by both traditional Fourier filtering and crystallographic image processing. Note that there are also more advanced Fourier filtering techniques which apply a Wiener-Kolmogoroff filter, but they require a reasonable model for the generalized noise [61] or at least for parts of it. In the absence of such models, one may just rely on the above-mentioned symmetrization on the basis of information theory and assume that the generalized noise is well approximated by a Gaussian distribution.

The work in [27-29] repeats the analyses of [26] and formulates interesting ideas on how further progress can be made (as of the years 2011/2013) in direct/physical space analyses of more or less 2D periodic images. Such progress is shown to be contingent on the utilization of *"adaptive classifiers"* as a possible (but partial) answer to generalized noise and pseudo-symmetry challenges. Such classifiers are non-binary/non-dichotomous in nature and result in the output of a list of reasonably well fitting plane symmetry groups. This list is ideally, but due to the existence of generalized noise not always, headed by the best fitting plane symmetry group.

References [27-29] define a twelve-dimensional vector with components that are translation symmetry normalized measures for the presence of characteristic symmetries in an analyzed image in direct space. The Euclidian distance of this vector to 22 twelve-dimensional reference vectors with binary components, i.e. 1 or 0, that represent perfect adherences to individual plane symmetry groups serves as a "symmetry distance measure" to gauge the closeness of the analyzed image to the 16 higher symmetric plane symmetry groups individually.

The binary reference vectors are referred to as symmetry "*prototypes*". The authors of [27-29] distinguish, however, between 23 different settings of the 17 plane symmetry groups, i.e. their symmetry prototypes, although there are only 21 such settings[9] in standard 2D crystallography [1-3,24,25].

[9] Plane symmetry group *p31m* and *p3m1* possess only one crystallographic setting each.

The approach of [27-29] proved to be superior to the computational dichotomous "yes or no" direct space approach of [26], which was referred to as an embodiment of a *"rule-based decision tree classifier"*. The dichotomous approach results in a classification of a noisy image into a single class [26], which is in [27-29] considered as a shortcoming since their classification output is a list of several reasonably well fitting classes that contains the best fitting class with a reasonably high probability. Whereas not explicitly stated in [27], these lists are in the best cases more or less a result of the naturally occurring translationengleiche supergroup-subgroup relations between the plane symmetry groups [2,3].

A step backwards with respect to what has been proposed in [26] is the utilization of normalized linear sums of residuals as components of the computed twelve-dimensional "symmetry representation vectors" in [27-29]. Such sums cannot straightforwardly become an integral part of a statistical model selection procedure that considers generalized noise as approximately Gaussian distributed.

Interestingly, the components of twelve-dimensional "symmetry representation vectors" in [27-29] as well as those of their eight-dimensional counterparts in [26] reveal the existing pseudo-symmetries in the analyzed images as byproducts in direct space. It is up to the researcher to decide which component of such vectors represents genuine symmetries that combine meaningfully to a plane symmetry group and which components represent only pseudo-symmetries. This is done in all of these papers by introducing some arbitrarily set thresholds into the analyses that may vary from image to image. Obviously, pseudo-symmetries can easily be mis-classified as genuine symmetries and vice versa when one needs to use such thresholds.

The most important idea in [27-29] is perhaps the suggestion to move conceptually away from *definitive* symmetry classifications. This author came to the same conclusion some time ago and well before he became aware of the corresponding idea in [27-29]. In contrast to [27-29], this author's approaches allow for quantifications of the probabilities with which a noisy 2D periodic image belongs to sets of crystallographic symmetry classification classes [22].

More or less 2D periodic images were analyzed in [26] without systematic considerations of the spatial relationships between translation symmetries and compatible point symmetries in direct space. In other words, arbitrary origin choices were assumed to be allowed for the Bravais lattices of all plane symmetry groups.

This is in contrast to the practices of standard 2D crystallography [1-3,24,25] where the origin[10] of a unit cell of a mathematical lattice defines the spatial distribution of all site symmetries in a mathematically consistent way. The authors of [26] proposed to the computational symmetry community, however, the usage of sums of squared residuals as components of their eight-dimensional symmetry representation vectors. This constituted significant progress in the year 2004.

[10] Sixteen of the 17 plane symmetry groups possess precisely defined (rather than arbitrary chosen) origins as illustrated graphically in [1,3]. These groups are referred to in this paper as the 16 higher symmetric plane symmetry groups. Crystallographic image processing relies on these 16 groups, whereas traditional Fourier filtering relies only on *p1*, which is the only group with an arbitrary origin choice in standard 2D crystallography.

The results of crystallographic symmetry classifications in direct/physical space are used to group images into classes of more or less 2D periodic images in openly accessible databases of everyday objects and scenes [62] or Islamic building ornaments [63] and to organize the sequence of sections in monographs that combine aspects of mathematics, crystallography, and the visual arts [64-67] for education and entertainment purposes. There are also proprietary image databases in the wallpaper, textile, and decorative ceramic tiles manufacturing industries which are used to inform and inspire new designs of such commercial products. The image processing program in [30-34] (named "FECETEX") has specifically been designed to support the acquiring, processing, storing, cataloguing, and retrieving of digital images in such databases.

In a remarkable development away from accepting strict bounds that are set by discrete mathematically abstract restrictions [1-4], crystallographic and non-crystallographic point symmetries have in direct space been defined as *"continuous features"* [45] in the year 1995. The "symmetry distances" of real-world objects such as the positions of diffraction spots in a Laue photograph of a quasicrystal[11] and mathematical objects such as irregular polygons to their nearest mathematically abstract symmetry representations, i.e. well fitting symmetry models, are in [45] expressed by minimized sums of squared residuals. There can be several initial symmetry models for each object that is to be classified and it is the task of the analyst to select the apparently best fitting model in the set. Sums of squared residuals are helpful but not sufficient to accomplish this task in a mathematically consistent manner when inclusion relations exist.

Kenichi Kanatani, i.e. the originator of geometric Akaike Information Criteria (G-AICs) [38-42], commented as early as 1997 on the approach of [45] and stated succinctly that it is for fundamental reasons incapable of dealing with symmetry hierarchies [42]. For an *objective* model selection procedure in the presence of symmetry hierarchies, pseudo-symmetries, and generalized noise, one needs to correct for the model specific geometric bias of the corresponding sum of squared residuals, i.e. use his G-AICs or perhaps either his geometric Bayesian Information Criterion (G-BIC) [68] or his geometric Minimal Descriptive (code) Length (G-MDL) [39] alternatives. Note that Kanatani's G-BIC and G-MDL are identical to each other as far as the final formulations of the criteria are concerned, but their derivations and internal logics differ.

The author's approach to plane symmetry group classifications utilizes a G-AIC [17,21,22]. As already mentioned above, the sixth section of this paper will give plane symmetry classification results by this approach for an experimental scanning tunneling microscope image [15] from graphite [50]. As also mentioned above, we will comment on the usage of the G-BIC/G-MDL alternative to such a classification in the penultimate section of this paper.

[11] The recorded X-ray diffraction pattern in the Laue geometry of [45] cannot be from an ordinary crystal as stated in that paper because its point symmetry is $C_{10}$ in Schoenflies notation. It must be from a quasicrystal instead due to its non-crystallographic Laue symmetry.

Kanatani demonstrated in [42] that when a supergroup of a point group explains the distribution of points in a distorted plane figure reasonably well, all of its subgroups (and in turn their subgroups) will have smaller symmetry distances as defined in [45] or by any other definition of a pure symmetry distance. His group theoretic insight applies, therefore, also to subgroups of plane symmetry groups that may represent an analyzed image reasonably well.

All of their subgroups (and in turn their subgroups) will actually have smaller symmetry distances, i.e. fit the image data even better, but contain less structure information (in the information theoretic sense). The Euclidian distances between twelve-dimensional vectors in [27-29] are, obviously, a pure symmetry distance and no exception in this respect. Kanatani's geometric bias corrections amend *pure* symmetry distances that are sums of squared residuals so that they can be used for model selection purposes in an information theory based sense.

Kanatani concluded more than 20 years ago that his kind of reasoning *"is expected to play a crucial role in building an intelligent system for automatically detecting a symmetry in an image"* [42], but was "ignored" by virtually everybody in the computational symmetry, Islamic building ornament appreciation, and applied crystallography communities for a long time. Credit goes to the authors of [26] to have floated the idea of using a G-AIC for the classification of more or less 2D periodic images into plane symmetry groups in the year 2004. As far as this author is aware, this has only been talked about in the computational symmetry community [32,35] and not actually been demonstrated there.

Typical for work in the computational symmetry community seems to be a focus on searching for the best possible algorithm to extract information from more or less 2D periodic images and to use that information to solve classification problems. The relative merit of different algorithms is typically assessed by classifying the same set of images from databases such as [62,63] and a comparison of the obtained classification accuracies.

Kanatani remarked in [39] that there are many algorithms for the extraction of geometric-structural features from images but that "*none of them"* is *"definitive"* in the sense that a single computer program will ever allow for a classification accuracy of 100 % for all possible images. What one should, therefore, be aiming for is a generalized noise level dependent classification accuracy that comes with a measure of how uncertain any obtained result is. The efforts of the computer scientists to develop the best possible and most accurate/precise algorithms for certain calculation and feature extraction tasks are in this context very much appreciated by the natural scientists.

A generalized noise level dependent classification accuracy is what natural scientists are in essence interested in. Any real-world measurement result in the natural sciences must be accompanied by an error estimate. The prevailing random noise during a real-world measurement limits the precision with which the result of the measurement can be known. Not being aware of some of the systematic errors and not accounting for them properly limits the accuracy with which the measurement results can be known. Algorithmic feature extraction from images for geometric-structural classification purposes is also a form of measurement so that natural scientists expect to be presented with both a classification result and an accompanying confidence level for that classification.

All of this author's approaches to crystallographic symmetry classifications are based on Kanatani's great work [38-42] and novel only insofar as they work in Fourier space and go partly beyond [21,22] his original proposals. This author's approaches [17,18] are characterized by the utilization of ratios of G-AICs, confidence levels of crystallographic symmetry classifications for symmetry models within the crystallographic symmetry hierarchy branch that possesses the highest likelihood, as well as individual Akaike weights for symmetry models that are disjoint [69] and Akaike weight products of complementary classifications [70] in cases of severe pseudo-symmetry challenges (of the Fedorov type [43]).

These classifications are *objective* because they are based solely on pertinent information in the images and the fulfillment of a series of numerically derived inequalities. These inequalities are either fulfilled or not, allowing for a statistically sound decision if there is sufficient structural information to conclude that a certain 2D Bravais lattice type, Laue class, or plane symmetry group is either present or not. The confidence levels of the classification results are obtained numerically from ratios of G-AICs for models within the same symmetry hierarchy branch [17,18].

All deviations from perfect periodicity and point symmetries in the images are in the author's approaches considered to be due to the unavoidable existence of image recording and processing noise as well as possibly existing structural defects in crystalline samples. Generalized noise includes all effects of (unavoidably) imperfect recordings of images, all kinds of rounding effects and numerical approximations by any kind of image processing algorithm, and all structural defects in crystalline real-world samples.

When there are many sources of this kind of noise and the effects of none of these sources dominate, the resulting generalized noise is by the central limit theorem approximately Gaussian distributed. This distribution is the precondition for the application of Kanatani's G-AICs, which are in essence geometric bias corrected sums of squared residuals. For the selection of the best fitting symmetry model within any one symmetry hierarchy branch, one does not need to estimate the generalized noise level as it can be eliminated in Kanatani's framework by algebraic means [17,18,22].

Conditional symmetry model probabilities within user-selected model sets [21,22] can also be calculated on the basis of these G-AICs and are particularly useful in cases of high levels of generalized noise in conjunction with Fedorov type pseudo-symmetries [43] so that it becomes *visually* impossible to distinguish between real symmetries and pseudo-symmetries. When one deals with disjoint symmetry models, on the other hand, one needs to estimate the generalized noise level. This is straightforwardly done by a "boot strapping" approach [17,18,22].

Crystallographic symmetry classifications by the author's approaches will always be generalized noise level dependent, i.e. somewhat preliminary in other words. This means that with better controlled experimental recording conditions and instruments as well as with better image processing algorithms and more perfect crystalline materials as samples in real-world imaging experiments in the future, the crystallographic symmetry classifications will become more precise. At any point in time, the classifications will have a numerically derived confidence level or conditional model probability as a measure of the (probabilistic) adherence of the image data to the classification class.

The author's information theory based approaches deliver with necessity only probabilistic crystallographic (or non-crystallographic) symmetry classifications because it is fundamentally unsound to assign abstract mathematical concepts such as a single 2D Bravais lattice type, a single crystallographic (or non-crystallographic) 2D Laue class, or a single plane symmetry group with 100 % certainty to a real-world image of a regular array of molecules on a crystal surface, or a crystal surface, or a crystal (or a quasicrystal). It makes, on the other hand, a lot of sense to report as result of a real-world measurement of a symmetry in an image not only the probability of the identification of the highest symmetric group/class present, but also to provide the corresponding probabilities of several of its subgroups/subclasses.

In contrast to the members of the computational [26-35] and visual Islamic ornament describing [36,37] symmetry communities, most practitioners of 2D crystallography [7-25] are interested in the best possible representations of both the content of the average unit cell and particularly the content of its asymmetric unit [1-3,24,25]. Crystallographic symmetry classifications are in the latter community often done for the express purpose of supporting the extractions of these contents. Electron and scanning probe microscopists are often also interested in suppressing the noise in the experimental images that they recorded with atomic or molecular resolution from (exceedingly thin) transmitted crystals [7,11-14] or regular arrays of molecules [7-10,15,18-20] on crystal surfaces.

Examples of such a studies are [11,12], where aberration-corrected scanning transmission electron microscopes have been utilized. There were, however, minor inconsistencies in the plane symmetry origin selections in these two papers.

As part of the crystallographic averaging of experimental images of clathrates, plane symmetry group *p4mm* and its translationengleiche [3] subgroups were enforced in [11] for demonstration purposes whereas it should actually have been *p4gm* and the corresponding subgroups. This was largely due to a misinterpretation[12] of the (entirely correct) information that is in [1] provided for the [001] projection of space group $Pm\bar{3}n$. A highly informative book chapter on aberration-corrected scanning transmission electron microscopy of electron beam sensitive materials [12] demonstrates that correct origin choices

[12] The International Tables for Crystallography list plane symmetry group *p4mm* as [001] projection of space group $Pm\bar{3}n$ (# 223) when an origin shift of $0,{}^{1}/_{2},z$ is applied [1]. The authors of [11,12] ignored this origin shift, which results in plane symmetry group *p4gm* being the projected symmetry of a full unit cell of the investigated crystalline clathrates along the [001] direction. Plane symmetry group *p4gm* features systematic absences of the odd *h0* and *0k* reflections whereas *p4mm* does not [3]. Crystallographic averaging in *p4mm* instead of *p4gm* resulted in [11] in some "confused" images due to origin "incompatibilities" for some of the translationengleiche subgroups of *p4mm*.

are essential to the crystallographic processing of images in Fourier/reciprocal space.

Such choices are supported in a computationally efficient manner by what is referred to as "*origin refinement*" in [13]. Origin refinement relies on the minimization of weighted sums of linear Fourier coefficient phase and amplitude residuals. These residuals do not foster an unambiguous symmetry model selection process that is based on Kanatani's geometric version [38-42] of information theory. The selection of one of the better fitting models without knowing (in the information theoretic sense) which is actually the model that contains the most structural information, i.e. provides the statistically best separation between generalized noise and structure, is left to the subjective experience of the researcher. The approach of [13] is, thus, not *fully* objective as the person who makes the symmetry classification does not have the benefit of pertinent sums of squared residuals that are corrected for the geometric bias of the individual symmetry models to guide her or him.

The weighted sums of linear Fourier coefficient phase angle and amplitude residuals of [13] will typically identify the symmetry hierarchy branch that contains the plane symmetry group that provides the statistically best separation between generalized noise and structure. Whereas the Fourier coefficient amplitude residuals will be rather similar for plane symmetry groups within any symmetry hierarchy branch in typical cases, the Fourier coefficient phase angle residuals will be lowest for all plane symmetry groups in the correct symmetry hierarchy branch. There is, however, no fully objective way to identify the plane symmetry group that is best in the information theoretic sense amongst the other plane symmetry groups that are also part of the correctly identified symmetry hierarchy branch with the highest likelihood of containing that group.

In other words, [13] provides no objective criterion on how far one can, in a statistically sound manner, "go up" in the correctly identified symmetry hierarchy branch with the highest likelihood in order to arrive at the crystallographic symmetry type, class, or group that provides the best separation of generalized noise from structural information. Going up in that branch means starting at the lower symmetric model and proceeding to the higher symmetric model as far as the fulfillment of the inequalities that involve ratios of G-AICs allows. The lower symmetric model will constitute a maximal subgroup or subclass of the higher symmetric model [2,3]. The higher symmetric model will in turn be a minimal supergroup or superclass of the lower symmetric model. In addition, the possible existence of pseudo-symmetries of the Fedorov type [43] as well as metric and motif based pseudo-symmetries [1,44] in the images to be classified remain as challenges to the standard 2D image based electron crystallography approach of [13,14].

On occasions, especially when there are pseudo-symmetries of the Fedorov type as well as high levels of generalized noise present in an experimental image, the subjectivity of the procedure in [13] will lead to structure refinements in the wrong plane symmetry group and subsequently to wrong structure reports in the literature. In these cases, the performed crystallographic image processing procedure that preceded the extraction of the atomic coordinates [13] will *not* be meaningful because it leads to wrong conclusions.

On other occasions of employing somewhat subjective crystallographic symmetry classifications in reciprocal space [13], the associated crystallographic image processing procedure may lead to results that are not as meaningful as they could be on the basis of the available experimental data. The subsequent extraction of atomic coordinates and structure refinement may in these cases have proceeded in a subgroup of the plane symmetry group to which the experimental image is actually closest in the first-order information theoretic sense. The refined positions of atoms or molecules will then not be completely wrong, just not as symmetrically distributed throughout the derived unit cell of the image as they are distributed (with maximized likelihood) in the crystalline sample [13]. The derived asymmetric unit will then be larger than it actually is when judged on the basis of the author's information theory based approach to plane symmetry group classifications [17,21,22].

The intensities of all of the pixels in all of the selected periodic repeats in a noisy 2D periodic image are in 2D electron crystallography averaged in reciprocal space in a manner that takes the various crystallographic restrictions of symmetrized models of the raw image data into account [13]. Geometric AICs are in the improved classification procedure used to *objectively* identify either the best symmetry model and give the confidence level of that identification [16,17] or to provide conditional model probabilities in the form of Akaike weights for a set of useful symmetry models either within a single crystallographic symmetry hierarchy branch or within competing branches [21,22] in cases of the presence of severe pseudo-symmetries of the Fedorov type [43].

Crystallographers work typically in 2D with images that are quite noisy due to their microscopic origin. This is particularly true when crystalline samples are investigated that can only tolerate a low dose of transmitted electrons [11,12]. Structural defects that are present in a crystalline sample are bound to be part of these images. As already mentioned above, the author considers structural defects in a sample as nothing else but contributors to the generalized noise in an image that has been recorded from that sample. These defects are typically not of interest to the 2D crystallographer as it is the average (ideal) structure of a crystalline sample that one wants to derive from an experimentally obtained image with atomic or molecular resolution [13].

Table 1 summarizes characteristics of the various groups of discussed analytical approaches to symmetry classifications of more or less 2D-periodic patterns, images, and geometric figures in 2D. This table deals mainly with classifications of the crystallographic translation and plane symmetries [7-37,44], but crystallographic and non-crystallographic point symmetries [42,45] are also mentioned.

TABLE I
Characteristics of the Analytical Approaches to Crystallographic (and non-crystallographic) Symmetry Classifications.

| | **Fully objective, i.e. researcher independent?** | **Definitive or probabilistic?** | **Generalized noise level dependent?** | **Applicable when only one (full) or a few unit cells are available?** | **Classification space?** | **References** | **Comments** |
|---|---|---|---|---|---|---|---|
| **Multiple authors on classifications by means of visual inspection and application of decisions trees** | no | definitive, but prone to misclassifications due to arbitrarily set thresholds and subjectivity | no | yes | direct | [24,25] and [36,37] | done in essentially the same way (without digital images and computers) for more than 100 years |
| **Liu and co-workers** | no | definitive, but prone to misclassifications due to arbitrarily set thresholds and subjectivity | no | yes | direct | [26] | decision trees within computer algorithms mimicking decisions that human beings would make, state of the art in 2004 |
| **Multiple authors in a comprehensive review and from the Polytechnic University of Valencia (Spain)** | no | definitive, but prone to misclassifications due to arbitrarily set thresholds and subjectivity | no | yes | direct | [30-35] | state of the art in direct space classifications up to 2009 |
| **Multiple authors from the Polytechnic University of Valencia (Spain)** | no | neither fully definitive nor fully probabilistic (in a quantitative manner),<br>misclassifications possible due to arbitrarily set thresholds and subjectivity | no | yes | direct | [27-29] | alternatives to applications of decision trees,<br>it was noted in 2011-2013 that aiming for a definitive classification is problematic because there are inclusion relations between many plane symmetry groups |
| **Traditional crystallographic image processing in transmission electron microscopy by other authors and this author's earlier work on generalizing that approach to other types of microscopies** | no | definitive, but prone to misclassifications due to arbitrarily set thresholds and subjectivity | no | no | reciprocal/<br>Fourier | [7-14] and [19-20] | very well established techniques, used for more than 50 years<br>Sir Aaron Klug's Nobel Prize in Chemistry (1982) was a recognition of both the importance of his development of crystallographic image processing techniques and the establishment of the field of structural biology on their basis. |
| **Zabrodsky and co-workers** | no | definitive, but no geometric bias corrections so that sub-figures are always preferred | no | not applicable as being concerned with 2D point symmetries of plane figures only | Direct | [45] | conceptual break-through in 1995 by considering symmetry as a continuous feature |
| **Kanatani in his comment on the work by Zabrodsky and co-workers** | yes | definitive, geometric bias corrections, without arbitrarily set thresholds | yes | not applicable as being concerned with 2D point symmetries of plane figures only | Direct | [42] | conceptual break-through in 1997 by bringing information theory to bear on considerations of hierarchic symmetries that are considered as continuous features |
| **This author's more recent work** | yes | probabilistic (in a quantitative manner), geometric bias corrections, without arbitrarily set thresholds | yes | yes when one works with larger images that are stitched together from the few available repeats | reciprocal/<br>Fourier | [15-18] and [21-23] | Applications of Kanatani's G-AIC for non-disjoint models. Application of Akaike weights for disjoint models. Systematic considerations of pseudo-symmetries of the Fedorov type. |

## III. Pseudo-symmetries of the Fedorov Type

Fedorov type pseudo-symmetries [43] are characterized by the fact that certain approximate plane symmetries are laid out on a crystallographic lattice. This lattice can either be the Bravais lattice of the plane symmetries that the hypothetical noise-free version of an image would possess or one of the superlattices of that particular Bravais lattice [23]. Genuine symmetries may then be mistaken for pseudo-symmetries and vice versa at both the metric and motif level, especially when high levels of generalized noise are present. Genuine symmetries and pseudo-symmetries "combine" to apparent "pseudo-symmetry groups" when the pseudo-symmetries are of the Fedorov type.

The images in Figs. 1 and 2 are synthetic and in open access [22,44]. On the left hand side of Fig. 1, there is the noise-free (original) version of the pair of images. The image on the right hand side of Fig. 1 has been obtained by adding independent Gaussian noise of mean zero and a standard deviation of 10 % of the maximal image intensity to the individual pixels of the noise-free image to the left.

One of the author's information theory based approaches to crystallographic symmetry classifications [17,21,22] allows for the identification of the correct plane symmetry group of the noisy image on the right hand side of Fig. 1 [23]. This image's subsequent symmetrization in this group leads to the most meaningful separation of structural information from generalized noise at the asymmetric unit level as shown in the upper inset of that image. The inset at the top of this image in Fig. 1 shows a single magnified unit cell as proof of the accomplishment of this separation of structure from noise.

Both images in Fig. 1 feature a Fedorov type combination of a motif-based pseudo-symmetry (sets of pseudo-four- and pseudo-two-fold rotation points plus vertical pseudo-mirror lines) and a translational (metric-based) pseudo-symmetry. One can, therefore, speak of a "pseudo-*p4mm*-symmetry" on top of the underlying *p1m1* symmetry (in long Hermann-Mauguin notation [1-4]). This is revealed in more detail (in direct space) in the reproduction of a magnified single unit cell of the noise-free version of the image as shown in Fig. 2a.

In reciprocal space, see both insets at the bottom of Fig. 1, the 01, 02, 04, and 05 spots are much weaker than the 03 and 06 spots as a result of the prevailing Fedorov type pseudo-symmetry. An "apparent periodicity" of strong spots in the amplitude maps of the discrete Fourier transforms of both images in Fig. 1 is revealed by the 03 and 06 spots (marked by white arrows in the inset on the left hand side) and in crystallographic standard terms [1] referred to as pseudo-periodicity. This pseudo-periodicity is a consequence of the prevailing pseudo-symmetry of the Fedorov type.

When moderate (to high [23]) generalized noise is added to the noise-free image on the left hand side of Fig. 1, spots that are already very weak (such as the 01, 02, 04 and 05 spots) "smear out" and tend to get "buried" in the noise so that their existence may be easily overlooked. As a consequence, one would then erroneously assign a pseudo-square reciprocal lattice to the inset at the bottom of the image on the right hand side of Fig. 1.

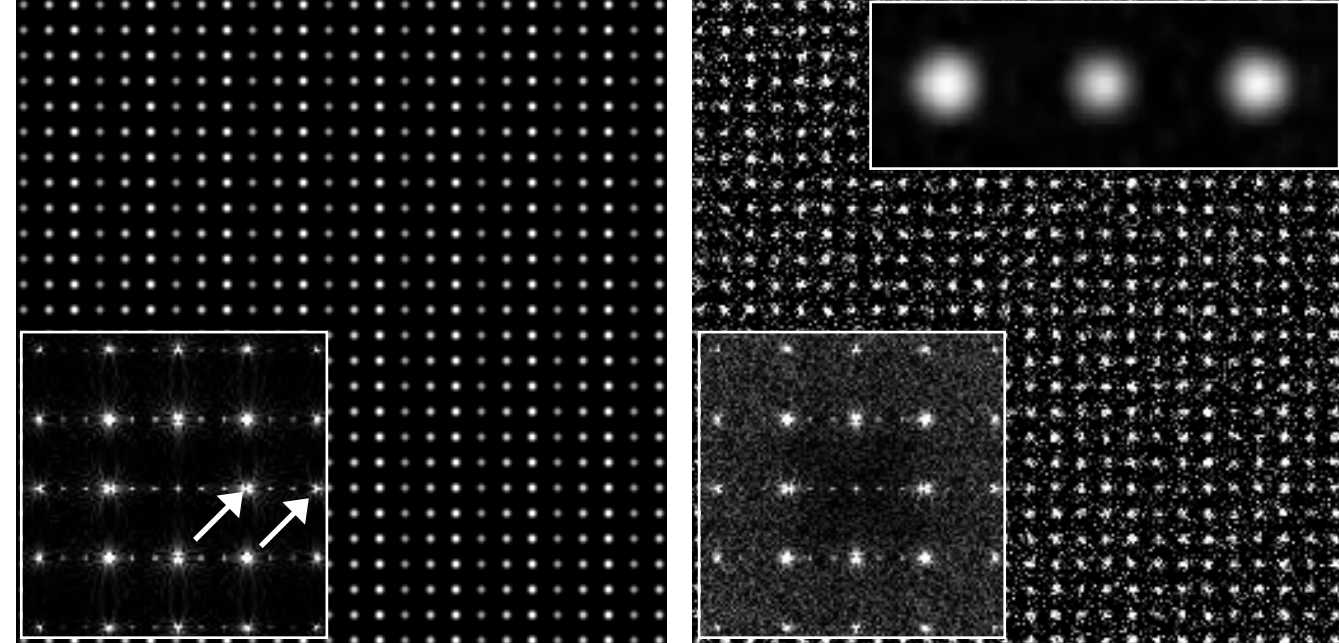

Fig. 1. Strongly pseudo-symmetric images with plane symmetry group *p1m1* on a rectangular Bravais lattice; left without noise, right with added noise. These images were created by Dr. Niklas Mevenkamp and sent to this author directly by Mevenkamp's PhD thesis supervisor Prof. Dr. Benjamin Berkels of the Institute for Advanced Study in Computational Engineering Sciences of the RWTH Aachen (Germany). Both images are reproduced from [44] with permission. The insets at the bottom of both images are the central parts of the amplitude maps (power spectra) of discrete Fourier transforms of the image intensity, as taken from [23] with permission. The 03 and 06 spots are marked by arrows in the inset on the left hand side. The additional inset at the top of the image with added noise is a magnified unit cell after symmetrization to the correct plane symmetry group. The asymmetric unit comprises either the upper or the lower half of this unit cell.

Two different computer programs that are dedicated to electron microscopy and crystallography and work in reciprocal space were in their default settings actually unable to extract lattice parameter information from the noisy image on the right hand side of Fig. 1 that would allow for its correct crystallographic symmetry classification into the rectangular Bravais lattice type [44]. The reason for this is the weak 01 spot in the inset of that image, which these programs simply ignored. Both programs took the 03 spot erroneously as end point of the reciprocal $\vec{b}^*$ vector instead. That vector became then equal in magnitude to the reciprocal $\vec{a}^*$ vector. A pseudo-symmetric unit cell that is in direct space three times smaller than the real unit cell of the images in Fig. 1 was the final result of this oversight [44].

Figure 2 reveals the prevalent genuine symmetries and pseudo-symmetries visually at the single unit cell level in direct space. Whereas there are only genuine mirror lines in Fig. 2a, the right hand side of that figure reveals pseudo-four-fold and pseudo-two-fold rotation points as well as additional pseudo-mirror lines (dashed) in addition to the genuine mirror lines, see Fig. 2b. An origin shift was applied to the right hand side of Fig. 2 for enhanced clearness. Whereas the origin of plane symmetry group *p1m1* is restricted to lie somewhere on one of the genuine mirror lines, one of the pseudo-four-fold rotation points in Fig. 2b has to be chosen as origin of the unit cell representation in pseudo-plane symmetry group $p_{b/3}4mm$ (in Chuprunov notation [43]).

Most readers will probably agree that the right hand side of Fig. 1 possesses apparently four-fold pseudo-rotation points, vertical pseudo-mirror lines, a Bravais lattice of the pseudo-square type [44], and consequently plane pseudo-symmetry group $p_{b/3}4mm$, see Fig. 2b. Note that the genuine *p1m1* plane symmetry and Bravais lattice of the rectangular type are both "sub-categories" of their pseudo-symmetry counterparts so that pseudo-symmetries can be very easily mistaken for genuine symmetries and vice versa when the former are of the Fedorov type and generalized noise levels are moderate, as on the right hand side of Fig. 1, or high as in [23].

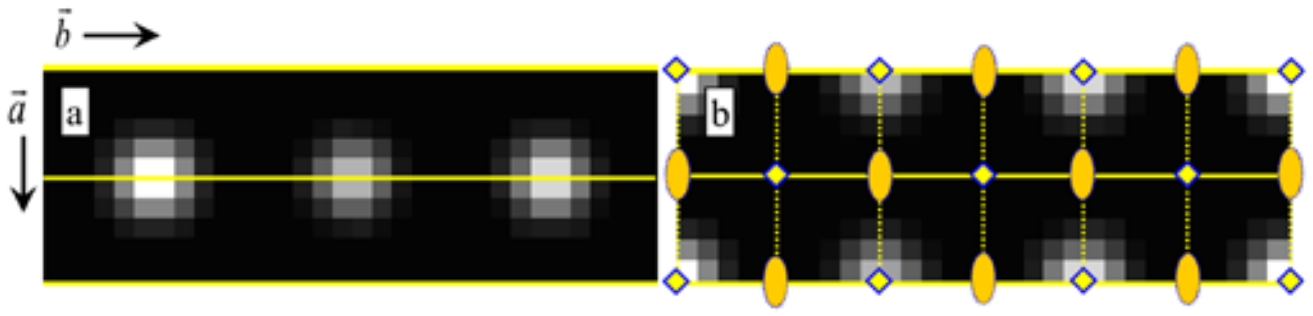

Fig. 2. Magnification of a single unit cell of the image on the left hand side of Fig. 1, i.e. the one without the added noise, from [22] with permission.
**(a)** The plane symmetry group is without any doubt *p1m1* (left hand side).
**(b)** Pseudo-symmetry group $p_{b/3}4mm$ [43] is made apparent at the unit cell level by marking the prevailing pseudo-symmetry operations (right hand side). An origin shift has been applied for added clarity. This origin shift is permissible in this plane symmetry group and does not change anything of standard crystallographic [1-4] relevance.

The quantitative results of the application of the author's plane symmetry group classification scheme to the two images in Fig. 1 (and a third one of the same series where the added Gaussian noise has a standard deviation of 50 % of the maximal image intensity) are given in [23]. The genuine crystallographic and prevailing pseudo-symmetries were clearly distinguishable in our studies (in spite of the high generalized noise level of the third image in this series [23,44] that is not shown here).

Note that one of the author's information theory based approaches to crystallographic symmetry classifications allows also for systematic "disentanglements" [21,22] of the effects of the genuine plane symmetries and possibly existing Fedorov type pseudo-symmetries. The disentanglement results will be generalized noise level dependent, but the noise level itself needs to be evaluated by a boot-strapping procedure [22] because different branches in the hierarchy of plane symmetry groups, Laue classes, and Bravais lattice types will be involved with necessity.

The relative Akaike weight [69] of each of the symmetry models in the selected set of models will give the probability that this model minimizes the expected Kullback-Leibler information loss [21,22]. For comprehensive crystallographic classifications in the presence of Fedorov type pseudo-symmetries, one can also calculate the products of the Akaike weights [70] for matching sets of symmetry models for plane symmetry groups, Laue classes, and Bravais lattice types [21,22].

## IV. Metric and Motif based Pseudo-Symmetries

A metric based (or translational) pseudo-symmetry exists in an image whenever there is a mismatch between the distribution of the site symmetries of a plane symmetry group and the apparent 2D Bravais lattice type. Figure 3 demonstrates such a mismatch on the basis of two synthetic images. A motif based pseudo-symmetry exists in an image when at least one genuine 2D site symmetry is apparently close to a higher site symmetry in the point symmetry hierarchy branch to which it belongs. Metric and motif based pseudo-symmetries can either co-exist in an image, as demonstrated in Fig. 3, or exist on their own.

The first of the two images in Fig. 3 is free of noise and given on the left hand side, whereas noise of the Gaussian type was added to that image with the freeware program GIMP to create the second image, which is shown on the right hand side of that figure. The author's information theory based approach to plane symmetry group classifications [17,21,22] allows again for the identification of the correct plane symmetry group of the noisy image on the right hand side of Fig. 3. Its subsequent symmetrization in this group leads again to the most meaningful separation of structural information from generalized noise at the asymmetric unit level as demonstrated in the upper inset of that image.

Noise in an image breaks all genuine symmetries and pseudo-symmetries alike. The differences between genuine symmetries and pseudo-symmetries at the motif level are, therefore, less visible in the image on the right hand side of Fig. 3. As many of the weakest spots in the amplitude map of the discrete Fourier transform get "smeared out" by the added noise and tend to get "buried" by it, see bottom inset of the image on the right hand side of Fig. 3, the relatively strong spots with $h \leq 3$ and $k \leq 3$ that possess pseudo-Laue class *4mm* begin to dominate the visual appearance of the image motif in both reciprocal and direct space.

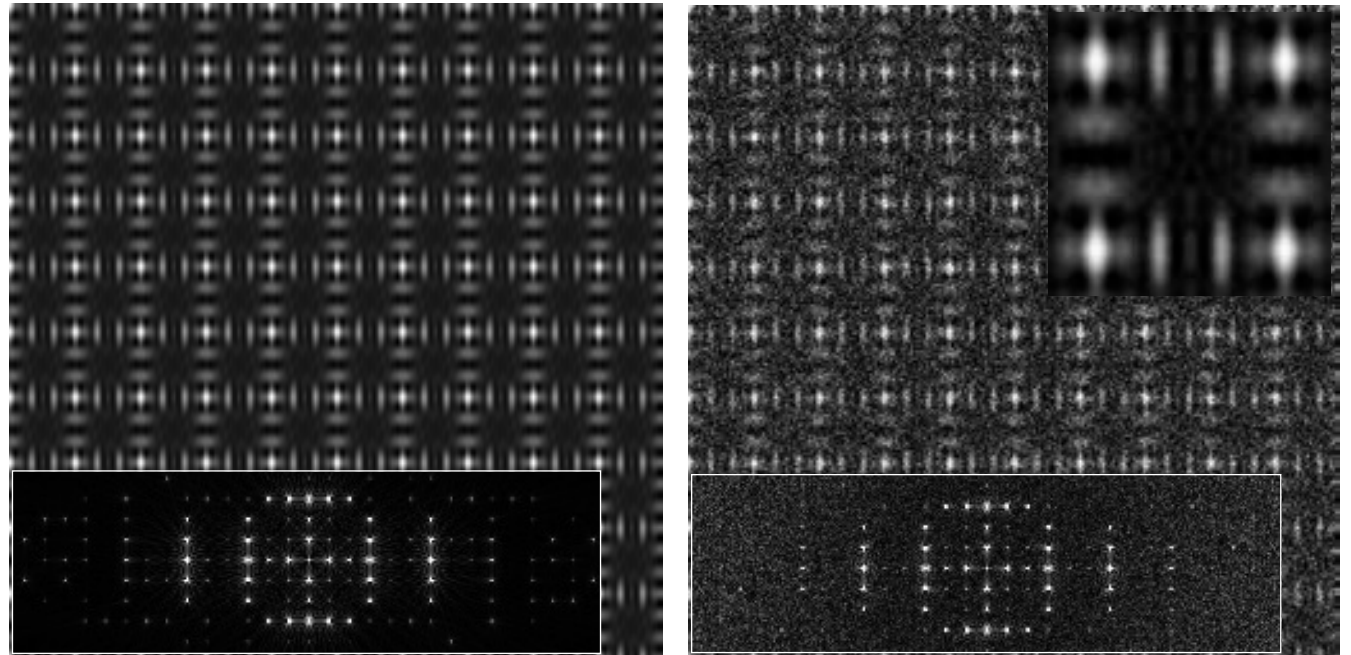

Fig. 3. Synthetic images with a combination of metric and motif based pseudo-symmetries as created by this author for the purposes of illustrating these two concepts and testing his information theory based approaches to crystallographic symmetry classifications. Whereas the image on the left hand side of the figure shows the noise-free version of the image, Gaussian type noise was added to the image on the right hand side. The insets at the bottom of both images are central sections of the amplitude maps (power spectra) of the discrete Fourier transforms of the image intensity. The additional inset at the top of the image with the added noise is a magnified unit cell (together with its immediate surroundings) after symmetrization in the correct plane symmetry group. The asymmetric unit comprises one quarter of this unit cell.

The lattice parameters of the two images on the left and right hand sides of Fig. 3 were extracted with the electron crystallography program CRISP [13,14] and are given in Table II.

TABLE II
Lattice Parameters of the Two Images in Fig. 3 as Extracted with the Electron Crystallography Program CRISP with Estimated Error Bars According to [44].

| Lattice parameters | $\lVert \vec{a} \rVert$ in pixels | $\lVert \vec{b} \rVert$ in pixels | $\gamma$ in degrees |
|---|---|---|---|
| Noise-free image | $51.9 \pm 0.5$ | $52.0 \pm 0.5$ | $90.1 \pm 0.5$ |
| Noisy image | $52.0 \pm 0.5$ | $51.9 \pm 0.5$ | $90.1 \pm 0.5$ |

Within typical error bars of the extraction of these lattice parameters [44], the corresponding Bravais lattice could be of the oblique, rectangular primitive, rectangular centered (when the lattice parameters refer to one lattice point), or square type. In crystallographic standard terms [1], an ambiguity such as this is said to be due to a metric specialization [44,71].

Some researchers would ponder that the assignment of the square Bravais lattice type would be justified for both of the images in Fig. 3. The (genuine) plane symmetry group is, however, per design *p2mm*, which does not contain site symmetries with a *4* in their symbol. There is, thus, an apparent mismatch as the prevailing *2mm* site symmetries require only a Bravais lattice of the rectangular primitive type [1,3]. We have, therefore, a case where the apparent translation symmetry would allow for the presence of site symmetries that contain a *4* in their symbol, but the genuine translation periodic motif contains no such symmetries.

In addition to the metric based pseudo-symmetry, there are sets of four-fold pseudo-rotation points associated with both images in Fig. 3 (as introduced per design). Motif based pseudo-symmetries are the consequence of the existence of these points. Such motif based pseudo-symmetries are always at the site symmetry [1,3] level and often exist independently of metric based pseudo-symmetries.

As consequence of the existence of sets of four-fold pseudo-rotation points and as a result of a casual inspection, genuine *2mm* site symmetries in the images in Fig. 3 may be mistaken for *4mm* pseudo-site symmetries, which would suggest the existence of plane symmetry group *p4mm* rather than *p2mm*. A more careful inspection of the left image in Fig. 3 reveals that the rotation symmetry of the pseudo-four-fold points is indeed only two-fold. Due to the added noise, it is more difficult to arrive at the same conclusion for the image on the right hand side of Fig. 3 by means of a visual inspection.

Due to the design of the images in Fig. 3, there is also an apparent "lattice centering pseudo-symmetry" that may lead one to the erroneous conclusion that the plane symmetry group of these two images might be *c2mm*. Since *p4mm* and *p4gm* are minimal non-isomorphic supergroups of *c2mm* [1-4], there is also a chance that one might erroneously assign one of these two groups to these two images and obtain a pseudo-unit cell that is twice as large as the genuine unit cell that is revealed in Table II. Because moderate to high generalized noise levels reduce the differences between genuine symmetries and pseudo-symmetries, noise levels and mis-classifications will be correlated.

Just as the reader will have had some difficulty in detecting the correct plane symmetry group in both images of Fig. 3 *visually*, most of the above and below discussed approaches to crystallographic symmetry classifications are also challenged by metric and motif based pseudo-symmetries. The author's information theory based approaches to crystallographic symmetry classifications overcome such challenges [72] as long as the first-order approximation in Kanatani's G-AIC is valid and its approximately Gaussian distributed generalized noise precondition is met. There are more examples of motif and metric based pseudo-symmetries for series of noise-free and noisy synthetic images in [44].

Note that it is for analytical approaches (other than the author's information theory based approach) typically more difficult to classify images such as the one on the right hand side of Fig. 3 with respect to its Bravais lattice type correctly due to the added noise (as demonstrated for other sets of synthetic images in [44]). The presence of noise does, on the other hand, not present challenges to the author's information theory based approach to crystallographic translation symmetry type classifications [18].

It is, however, quite conceivable that there are noisy 2D periodic images with extreme cases of motif and metric based pseudo-symmetries that even the author's information theory based approaches to crystallographic symmetry classifications would mis-classify (at some point in time) because their results are always generalized noise-level dependent at any one point in time. When such images are in the future recorded from the same samples with lower noise levels and subsequently processed with more precise algorithms, the author's approaches will deliver fewer and fewer mis-classifications.

## V. 2D Translation Symmetry Classification by a Neural Network and Comparison to 3D Results from other Neural Networks

A deep convolutional neural network (DCNN) has recently been employed [46] to the crystallographic symmetry classification of atomically resolved images from transmission electron microscopes and scanning tunneling microscopes into 2D Bravais lattice types. The authors of that study noted that not all 2D Bravais lattices types are disjoint, as it is often the case in more typical (non-crystallographic) image classification tasks that are performed by DCNNs, so that their task was somewhat more challenging than usual. They trained their network with a set of 24,000 computer generated images that contained 4,000 members for each of the five 2D Bravais lattice types, i.e. 20,000 more or less 2D periodic images, and an additional class of non-periodic images that included images without spatially resolved atoms. We refer here to these two different types of classes as "crystal images" that do possess 2D Bravais lattices and "non-crystal/empty images" that do not.

The members of the non-crystal/empty image class are in essence devoid of distinct periodicities in Fourier space. More precisely, they are devoid of well resolved peaks in the amplitude maps of discrete Fourier transforms, which would allow for their classifications into one of the five Bravais lattice types [44] that exist in the Euclidian plane. Perfectly 2D periodic crystal images of the training dataset were subjected to randomizations of the atomic positions. These randomizations can be taken as loosely equivalent to introducing generalized noise (that includes real-world image recording and processing noise) into the crystal images. The effect of these randomizations in direct space was some smearing out of the Fourier coefficient amplitude peaks in reciprocal space. When those peaks get smeared out too much, classifications as non-crystal/empty images should result and have indeed been observed [46].

Correctly labeled images with pseudo-symmetries of the Fedorov type [43] with well defined noise levels [44] were missing from the training set of images [50]. Such images are for example shown above in sections III and IV of this paper and in recent papers by this author [21-23,44]. The labeling of these kinds of training images should best be done in an *objective* manner by means of the application of G-AICs, as briefly mentioned above and demonstrated in [23].

This fact makes the present paper relevant for future training rounds of the above-mentioned crystallographic DCNN and for the development of more sophisticated neural networks that may eventually allow for comprehensive crystallographic symmetry classifications from noisy 2D periodic images in the presence of pseudo-symmetries of various types. The omission of objectively labeled pseudo-symmetric training images with varying noise levels means necessarily that the DCNN that is described in [46] must have difficulties in classifying corresponding experimental test images and synthetic validation images correctly into their respective 2D Bravais lattice types.

Two experimental scanning tunneling microscope (STM) images were also classified with respect to their translation symmetry type with the DCNN machine in [46]. Due to the nature[13] of the contrast in STM images, their plane symmetry is not necessarily the one which is obtained by an orthogonal projection from the 3D space group symmetry, as listed in [1] for several high symmetry directions. Graphite [50] served as sample in both of the classifications of experimental STM images in [46].

One of their two samples had been drifting during the recording of its STM image and the DCNN machine classified the translation symmetry of the corresponding image as being of the oblique Bravais lattice type [46]. This result is as it should be because crystallographic symmetries get severely broken when a sample does not remain stationary during the recording of its image. For their stationary graphite[14] sample, they obtained the hexagonal Bravais lattice type [46].

[13] It is well known that the local electronic density of state at the Fermi level is responsible for the contrast in STM images. The symmetry of the probed near-surface region is to be described by an orthogonal projection of a subperiodic (3D) layer.

[14] That STM image is freely downloadable (in the *.tif format) from the on-line support material of [46]. This author has, however, so far not reproduced this particular result on the basis of an information theory based Bravais lattice type classification. Plane symmetry group *c11m* was obtained in 2020 by G-AIC as the one that minimizes the Kullback-Leibler divergence of that image. This indicates that the Bravais lattice of this graphite sample could, in the absence of a metric specialization, be of the rectangular-centered type. An earlier analysis on the basis of a *.jpg version of this STM image [16] confirmed either preemptively or erroneously the hexagonal Bravais lattice type for this image.
The symmetrized version of that experimental STM image in plane symmetry group *p31m* revealed both carbon atoms in the unit cell [16,73]. Whereas other researches have observed both carbon atoms in primitive unit cells of graphite in raw and processed experimental STM images over the years before, many more such images revealed only one such atom [74]. The physical causes for the very pronounced differences in the STM imaging contrast of these two atoms seem to be well understood [75].
Note that plane symmetry group *h3m1* is laid out on a triple hexagonal cell that reduces to *p31m* after re-indexing to a primitive unit cell [1]. The relatively low sum of squared residuals of the complex Fourier coefficients for these two symmetries is highly suggestive of a pseudo-rhombohedral symmetry for the electronic density of state in this particular graphite sample, at least in the local vicinity of the conspicuously visible structural defects that reduce a presumably genuine three-fold symmetry to a pseudo-symmetry.

Of the three papers that used DCNNs and synthetic training datasets in reciprocal space [46-48] without including pseudo-symmetric images (with or without significant noise levels), the classification accuracies in the latter two of these studies are superior to that of the former. This might possibly be due to the three-dimensional nature of the symmetry classifications in the latter two papers [47,48] as discussed in the appendix.

Particularly impressive is that the DCNN of [48] dealt well with metric specializations and extreme cases of translational (metric-based) pseudo-symmetry in 3D whereas the DCNN of the 2D study of [46] failed in this task completely (as demonstrated by supplementary figure 5 in the on-line supporting material to [46]). Note that the classifications of [48] were restricted to a few classes with very high 3D symmetries. General trends may be derived from the comparisons of the DCNN classification results of [46-48]. These trends seem to be: the more remnants of genuine symmetries there are in noisy data, the easier the classification, and the more accurate the classification results.

There were incorrect usages of crystallographic standard terms for the 2D Bravais lattice types [1,3,44] as well as misleading statements about the nature of some of their samples [50], the existence of clearly discernable Moiré fringes in the amplitude map of the discrete Fourier transform of one of their images [16], and a figure which implied that classification probabilities by their DCNN could be negative or in excess of 100 % in the original version of the paper by Vasudevan and co-workers [46]. The crystallographic misnomers have been corrected to a large extent in the current (December 18, 2018) on-line version of their paper, i.e. almost six months after the original publication.

## VI. Geometric Bayesian Information and Minimal Descriptive length Criteria as Possible Alternatives to geometric Akaike Information Criteria

It needs to be acknowledged that there are other geometric information criteria such as Kanatani's geometric Bayesian Information Criterion (G-BIC) [68] and his geometric Minimal Descriptive (code) Length Criterion (G-MDL) [39] that could be utilized for crystallographic symmetry classifications instead of Kanatani's geometric Akaike Criteria. As noted above, Kanatani's G-BIC and G-MDL are identical as far as their formulations are concerned, but their derivations and internal logics differ.

It is also well known that G-AICs tend to select models that are *"faithful to the data"* [68] so that erroneous selections of higher symmetric models, i.e. crystallographic supergroups [1-4] over lower symmetric models, i.e. crystallographic subgroups [1-4], are rare. Kanatani's G-BICs and G-MDLs prefer, on the other hand, models with less complexity due to the larger geometric bias correction term.

Unpublished simulations based on synthetic images with strong pseudo-symmetries have in 2020 shown that the G-AIC model selection process works for plane symmetry groups and Laue classes very well as long as noise levels are zero, small, or medium. Such model selections may, however, fail for high and very high noise levels due to a tendency of strong pseudo-symmetries being mistaken for genuine symmetries when noise levels are high. The resulting symmetry classifications were in such cases in favor of higher symmetric groups and classes.

## VII. Summary and Conclusions

The existing types of classification approaches for the crystallographic symmetries of more or less 2D periodic images in direct/physical and reciprocal/Fourier space were discussed and their relative performance evaluated in a qualitative manner. A few details on the author's information theory based approaches were given as they allow for (fully) objective classifications of crystallographic symmetries such as Bravais lattice types, Laue classes, and plane symmetry groups of real-world images that are more or less periodic in 2D. It was also demonstrated that the author's approach to plane symmetry group classifications works well for synthetic but noisy and severely pseudo-symmetric images.

Whereas at least some of the deviations from perfect periodicity are due to the unavoidable existence of image recording and processing noise, possibly existing structural defects in crystalline samples contribute to the generalized noise level on an equal footing. Information theory based classifications into plane symmetry groups enable the best possible separation of structural information from generalized noise and provide, thus, the basis for the most meaningful crystallographic averaging in the spatial frequency domain.

This kind of averaging is over all processed asymmetric units, whereas traditional Fourier filtering averages only over the processed unit cells. As the area of the asymmetric unit is up to 12 times smaller than the unit cell, the former type of averaging suppresses generalized noise much more effectively than traditional Fourier filtering. In effect all deviations from a prescribed crystallographic symmetry are removed from an image by this form of image processing so that the averaged structure emerges clearly whereas it is only the average translation symmetry that is enforced by traditional Fourier filtering.

Pseudo-symmetries of different kinds were discussed because they present severe challenges to most crystallographic symmetry classification schemes when high levels of generalized noise exist in the images. As one would expect, the author's approaches to crystallographic symmetry classifications overcome these challenges as long as the first-order approximation in Kanatani's G-AIC is valid and its approximately Gaussian distributed generalized noise precondition is met.

Because it is fundamentally unsound to assign an abstract mathematical concept such as a single symmetry type, class, or group with 100 % certainty to a more or less 2D periodic record of a noisy real-world imaging experiment that involved a real-world sample, the author's approaches to crystallographic symmetry classifications deliver with necessity only probabilistic classifications.

Recent applications of deep convolutional neural networks to crystallographic symmetry classifications in two (and three) dimensions by other authors were discussed (in an appendix) as they deliver probabilistic classifications by other, i.e. *non-analytical*, means. The discussed "correlation detection and optimization" approaches remain (at least so far) ignorant of the fact that many crystallographic symmetries are actually hierarchic, i.e. that the classification classes are often not disjoint.

Both the analytical and the machine-based/non-analytical probabilistic approaches can be divided into those which work in direct/physical space and those which work in Fourier/reciprocal space. Between all of the discussed approaches, the ones by this author are currently unique insofar as they do not involve arbitrarily set thresholds and subjective judgments. The corresponding classifications are dependent on the generalized noise level and will, therefore, always have a somewhat preliminary character at any one time. Nevertheless, they are quantitative as confidence levels and/or conditional model probabilities are provided.

Improved imaging techniques and image processing algorithms with fewer approximations and larger numerical precision will in the future lead to reduced generalized noise levels for images that were taken from the same samples. Crystallographic symmetry classifications of more or less 2D periodic images with the authors information theory based approaches will, therefore, tend to obtain higher confidence levels (or model probabilities) of correct classifications in the future.

## Appendix: Classifications of 3D Crystals by Neural Networks and Associated Background

Deep convolutional neural networks were recently also applied to synthetic "images" that represent crystallographic symmetry information in 3D [47,48]. Just as in the case of the neural network for 2D Bravais lattice type classifications [46] (as discussed in the fifth section of this paper), pseudo-symmetries present challenges to these networks in conjunction with generalized crystal structure data recording and processing noise [44], but were neglected in both of these studies. Because noisy images with G-AIC based labels were not part of the training sets of these neural networks that worked with crystallographic 3D data [47,48] (as judged from the available information in these papers), the classification performance of these networks must also be limited.

We will discuss the earlier of these two papers first. The "images" of that study [47] consisted of the calculated powder X-ray diffraction patterns of approximately 150,000 entries of the Inorganic Crystal Structure Database (ICSD) [51]. The crystallographic symmetry classifications were done with respect to the seven crystal systems, 101 extinction symbols (rather than extinction groups[15]), and 230 space group types that exist in 3D so that one can meaningfully claim that this network is composed of three distinct DCNNs that are working on complementary aspects of crystallographic symmetry classifications. Note that extinction symbols represent partial space group information, but it is incorrect to refer to these symbols as "extinction groups" as it has been done by Park and co-workers in their paper. Whereas Poisson noise was added to the calculated crystal structure information bearing images, effects of crystallite textures[16] were ignored [47].

[15] The internationally accepted standard reference of crystallography in two and three dimensions [1] does not use the concept of "extinction groups" as employed in [47]. Extinction symbols are used instead and represent partial space group information that can be derived from the fulfillment of the reflection conditions. All extinction symbols start with a capital letter for the prevailing 3D Bravais lattice type that is followed by up to three of the subsequent lower case letters of the full Hermann-Mauguin symbol of a space group type replaced by a dash (-). For example, *Pn--* is the extinction symbol for space groups $Pnm2_1$ and *Pnmm*. The terms extinction symbol and diffraction symbol are used more or less synonymously in the crystallographic literature.

Eighty percent of the calculated images were utilized in the training dataset and the remaining 30,000 images served as the validation dataset. A total of 10,001 input "neurons" represented all of the calculated powder X-ray diffraction intensity values in the 2Θ range between 10 and 110°, i.e. the so-called total (or full) X-ray diffraction profile. Classification accuracies[17] of 81.14, 83.83, and 94.99 % were achieved after the training process for the space-group, extinction-symbol, and crystal-system classifications, respectively. Unsurprisingly, these validation results showed that the larger the number of possible crystallographic symmetry classes in a DCNN, the lower the obtained classification accuracies [47]. Given the fact that there are a total of 230 output neurons for the space group symmetry classification, the obtained 81.14 % classification probability is quite impressive.

Two experimental X-ray diffraction patterns/"images" that (*i*) the network had never encountered before, (*ii*) that did not belong to any of the 9,093 structural prototypes of the ICSD, and (*iii*) which contained small amounts of crystalline impurities were used as test cases. The crystal system was correctly determined in both of these test cases, but the obtained space group and extinction symbol classifications were wrong. The latter result may be due to the facts that (*i*) the two experimental test datasets did not belong to any of the known structural prototypes of the ICSD and, (*ii*) the experimental samples contained small amounts of crystalline impurities, as it is often the case in experimental studies.

At the present time, this author agrees wholeheartedly with the closing statement of the paper by Park and co-workers: *"This small success will be a milestone for the further development of deep-learning-based analysis for many other conventional theoretical rule-based tasks in materials science."* [47]. A comment is in order here on the differences in the semantics of the word "rule" as used by members of the computer science community (as in the preceding sentence), on the one hand, and members of natural science communities, on the other hand.

Some members of the natural science communities may prefer to characterize their endeavors as applying "physical laws" rather than "rules". When they are working with empirical data that are contaminated by noise and the natural scientists do not employ inferences on the basis of information theory, e.g. traditional (frequentist) Akaike Information Criteria [76], G-AICs [38-42], conceptually similar measures [68], Akaike weights [69], and their products [70] for the selection of one mathematical model for the data over another based on multiple working hypotheses [77], the natural scientists employ "physical laws" necessarily in a somewhat subjective manner so that mathematical abstract laws convert unavoidably to personal rules and preferences. Natural scientist may not always be fully aware of this fact. Computer scientists are, on the other hand, generally aware of this fact and this leads them to use the word "*rule*" when they want to describe what experts in any science field employ.

What was not explicitly discussed in [47] is the fact that the result of a crystallographic symmetry classification will only be as good as the training dataset allows. Whereas this is perfectly obvious to any computer scientist, the reasons why there are problems with the currently obtainable training datasets need to be elaborated a little bit.

As discussed in the appendices of [22], it is well known that a few percent of the entries in the ICSD are either misleading or wrong. Major reasons for this are unrecognized pseudo-symmetries of the Fedorov type, metric and motif based pseudo-symmetries, and subjectivity in employing rules to arrive at the crystallographic records of experimental X-ray crystallography studies that ended up in the highly venerable Inorganic Crystal Structure Database. The conclusion must, therefore, be drawn that crystallographic symmetry classifications by DCNNs that rely on the current state of the art in mainstream 3D crystallography are limited by the correctness of the labels on the training data. The future application of information theory based approaches to crystallographic symmetry classifications in mainstream 3D crystallography is poised to correct that problem over time.

Three-dimensional crystallographic information was recently taken from the elemental sub-section of the AFLOW-LIB database [52] and converted into 3D structure information bearing "images" in the Euclidian plane [48]. Ziletti and co-workers created in effect a novel type of "diffraction fingerprint" for a large fraction of the structures into which many chemical elements were calculated to crystallize [48].

In essence, these fingerprints represent the superposition of six zone-axis diffraction pattern that are obtained by a crystal tilting protocol and are calculated for a wavelength that corresponds to a very large Ewald sphere. The intersections of this sphere with the reciprocal 3D crystal lattice for all six incident beam directions of the tilt protocol are thereby sufficiently well approximated by 2D planes. The diffraction information in the corresponding six crystallographic planes is combined into a single 2D "image" which contains genuine 3D structure information, but represents only a small fraction of the obtainable totality of that information.

[16] A crystallographic texture exists in a polycrystalline sample whenever there are preferred (rather than random) crystallite orientations with respect to some external coordinate system. Different types of crystallographic textures are frequently encountered in powder X-ray crystallography and are often caused by specific crystallite morphologies. They lead unavoidably to discrepancies between the experimentally recorded intensities of Bragg peaks and their counterparts that were calculated on the basis of the assumption of a completely random distribution of the crystallite orientations, the structure factors, and the multiplicity of the Bragg planes that contribute to a Bragg peak. The multiplicity of the Bragg planes is a feature of the prevailing space group symmetry.

[17] Note that the two digits after the decimal point in the classification accuracies are not significant numbers in a real-world scientific sense. They are just the natural outcome of dealing with a very large number of 3D crystal information bearing "images".

There are at present some 3,000 calculated entries in the AFLOW-LIB elemental database [52] and it is not clear how many of them were converted into structural fingerprints and served as the training and validation datasets in [48]. Only the ratio of training to validation data is known to be nine to one for the so called "*pristine dataset*", which contains only structures that could unambiguously be classified with respect to their symmetry by the computer program Spglib [53] when it was used with some reasonable threshold [48]. As an unavoidable byproduct of their data selection procedure using a subjectively set threshold, the pristine dataset cannot contain many noisy diffraction fingerprints of pseudo-symmetric structures so that their trained DCNN is with necessity unable to classify such structures correctly.

Approximately 250 atoms, comprising some 60 to 250 unit cells, were used in each case for the calculation of the diffraction fingerprints [48]. In order to account for structural crystal defects in a very general sense, atoms were randomly displaced, substituted or removed from the structures in the pristine dataset in order to create a much enlarged validation dataset. One may consider this kind of pseudo-real structure modeling as being somewhat equivalent to adding generalized "image" recording and processing noise that any experiment based diffraction fingerprint would possess.

Because all symmetries were broken in the validation dataset of the defective structures, weak spots appeared in many diffraction fingerprints because extinction rules were no longer obeyed due to the random removal and/or substitution of atoms of the pristine dataset. The DCNN was, however, perfectly capable of ignoring these weak spots [48] as it was trained only on the pristine dataset where the extinction rules are perfectly obeyed.

There are 81 stable (i.e. non-radioactive) elements in nature, but just eight space group symmetry types account according to [48] for more than 80 % of the space symmetries of their crystal structures. In the order of falling space group numbers as listed in the International Tables for Crystallography [1] these space group types are: $Im\overline{3}m$ (# 229), $Fd\overline{3}m$ (# 227), $Fm\overline{3}m$ (# 225), $Pm\overline{3}m$ (# 221), $P6_3/mmc$ (# 194), $R\overline{3}m$ (# 166), $I4_1/amd$ (# 141), and $I4/mmm$ (# 139). These eight space group types represent the totality of the possible classes for the crystallographic symmetry classification of the DCNN in [48].

All of the above-mentioned space group types are centrosymmetric just as all calculated diffraction fingerprints are centrosymmetric. Due to the particulars of the employed tilt protocol, the calculated diffraction fingerprints turned out to be identical for database entries with space group types # 141 and # 139 so that crystal structures with both space group types needed to be classified into a single class. Other tilt protocols are bound to reveal significant difference in the diffraction fingerprints that should allow for classifications into both space group type classes.

When referring to the crystal structures that many chemical elements take in solid form, several of the above listed space group types represent the symmetries of what is well known in materials science as structural prototypes [54]. It is, therefore, more correct to state that Ziletti and co-workers performed classifications for calculated pristine and defective element structures into a few elemental structural prototypes [48] rather than into a few (much more general) space group types[18].

Note that most of the above-mentioned space group types are highly symmetric and several of the classification classes are disjoint so that very high classification probabilities are to be expected on theoretical grounds for the DCNN in [48]. Unsurprisingly, very high classification probabilities were indeed obtained. If a wide range of non-disjoint space group symmetries, i.e. space group types within symmetry hierarchy branches, representing structural prototypes of a multitude of crystal structures of chemical compounds (rather than only elements) were involved in a similar DCNN study and Kanatani's deep group theoretic and statistics insights [38-42] had again been ignored, the classification probabilities would have conceivably been much worse.

Reference [48] states incorrectly that the authors of that paper made classifications into "crystal classes" while they were in reality making classifications into a few structural prototypes of crystalline element structures. Their structural prototypes are characterized by small selections of both possible translation symmetries and systematic reflection absences. The correct crystallographic meaning of crystal classes including the distinction between geometric and arithmetic crystal classes is given below in a footnote[19].

It is explicitly noted in [48] that the traditional approach to crystallographic space group symmetry classifications in 3D involves subjectively set thresholds. The method of [48] was erroneously presented as being free of such subjectivity and, therefore, allegedly superior to the traditional *analytical* approaches of mainstream 3D crystallography. Because the training dataset was constructed by attaching labels to calculated structural fingerprints on the basis of outputs of the Spglib program [53], the *arbitrarily* set thresholds of that computer program made sure that the DCNN based classification results in [48] must be somewhat subjective.

It is noteworthy that [48] states that traditional, i.e. non G-AIC based, *"approaches to space group determination fail in giving the correct (most similar) crystal class in the presence of defects"*. Structural defects are in a G-AIC based classification approach simply part of the generalized noise level so that the application of G-AICs is free from subjectivity. To this author's knowledge, there have, however, not been any G-AIC equations specifically derived for mainstream 3D crystallography so that it is no surprise that they have not yet been used by mainstream 3D crystallographers.

[18] A space group type is the space group symmetry of a multitude of structural prototypes. The ICSD distinguished for example between more than 9,000 inorganic structural prototypes [51] and a good college-level materials science textbook [54] lists 99 structural prototypes for 46 space group types explicitly. There are 230 space group symmetry types in total.

[19] There are two different types of crystal classes. The geometric crystal class refers to the symmetry of the external shape of macroscopic crystals. There are 32 geometric crystal classes in 3D and 10 such classes in 2D. Both of them have a one to one correspondence to the 32 and 10 point group symmetries in the Euclidian 3D space and 2D plane, respectively. An arithmetic crystal class, on the other hand, refers to a combination of a geometric crystal class with a Bravais lattice type. There are 73 arithmetic crystal classes in 3D and 13 in 2D.

The author of this paper agrees with the third anonymous peer reviewer that the scope of [48] *"is so narrow ... as to prove only the concept rather than demonstrating a real materials breakthrough"* (on-line supporting material to [48]). Experimental data from the atom probe tomography of metallic element crystals are probably well suited for classifications with the DCNN of [48] due to the facts that (*i*) up to approximately 20 % of the atoms may have escaped detection by current state of the art detectors and (*ii*) many metallic elements crystallize in a very small number of centrosymmetric structural prototypes.

An interesting feature of the discussed application of DCNNs to crystallographic symmetry classifications (in reciprocal space) is that the outputs, e.g. a particular crystallographic symmetry type or a structural prototype (that possesses a space group and Bravais lattice) come naturally with a measure of the probability of the assignment to that particular class or type. When the networks are configured so that multiple working hypotheses [77] are tested simultaneously, one obtains the conditional probabilities for the classification into a range of classes and types within the selected hypotheses set.

The classification probabilities of DCNNs are, however, only measures of the "correctness" of a classification of an unknown test sample when one can be sure that all of the labels on noisy training and validation data were correct to 100 %. As stated above, there cannot be such an assurance as long as information theory based approaches to crystallographic symmetry classifications are not implemented.

The last paper that is going to be mentioned in this appendix is about the application of two "off the shelf" DCNNs with pre-trained model weights to the classification of very low magnification images that were taken with a light microscope from representatives of 13 different crystalline compounds [49]. A total of approximately 7,000 images were recorded. Seventy-five percent of these images served as training set and the remaining 25 % as the validation set. A total of 180 images from one crystalline test sample, i.e. urea, were then classified with probabilities of 93.34 and 99.41 % by the two DCNNs, respectively.

The prevailing individual crystal morphologies, i.e. 3D geometric crystal classes and the isomorphic 3D point symmetry groups, will have contributed to these quite impressive test results, but were not discussed in [49]. It is well known that sufficiently developed faces and facets allow for the identification of crystalline materials by means of optical goniometry [55] in conjunction with associated databases [56,57]. The application of the optical goniometry technique is, however, rather time consuming and requires a high level of crystallographic background knowledge, manual skills, and special instrumentation that is seldom used nowadays.

Note that the above-mentioned classification probabilities refer to the identification of the test images as originating from urea crystals [49]. They are definitively not the probabilities of crystallographic symmetry classification. This fact makes that study very different from the other two studies that are discussed in this appendix and section five of the main body of this paper.

The study of Mungofa and coworkers [49] is, therefore, effectively free of the ambiguities that are inherent in crystallographic symmetry classifications by means of current state of the art DCNNs [46-48] that ignore naturally occurring inclusion relations [1-3,24,25], pseudo-symmetries, and Kanatani's information theoretic approaches [38-42] to deal with them objectively. The relatively simple study of [49] is, therefore, of a high practical value at the present time, while there are unresolved problems with the other three DCNN studies [46-48] that limit their value.

Overcoming these problems will take some time as they are rooted in the subjectivity of most practitioners of the current state of the art of mainstream 2D and 3D crystallography. With respect to the usage of neural networks within the natural sciences including crystallography, this author agrees (at the present time) with the conclusions of a recent trade journal article [58] that *"machine learning is overhyped, won't cure cancer ..."* but is *"... a valuable tool that's here to stay"*.

## Acknowledgments

Professor Bryant York of Portland State University's Computer Science Department as well as Andrew Dempsey and Connor Shu of the author's research group are thanked for careful proof readings of various drafts of the manuscript. The latter two persons are also thanked for both (*i*) developing MATLAB code that supports crystallographic symmetry classifications of noisy 2D periodic images according to the author's information theory based approaches and (*ii*) delivering intermediate calculation results that contributed to the crystallographic classifications of the two sets of synthetic images in this paper and of one of the experimental scanning tunneling microscope image that was put into open access by Dr. Rama Vasudevan of the Oak Ridge National Laboratory. Doctor Rama Vasudevan and his colleague Dr. Maxim Ziatdinov (of the Oak Ridge National Laboratory) are thanked for private communications. Professor Manuel Contero of the Polytechnic University of Valencia (in Spain) is thanked for a private communication, for uploading one of his papers, i.e. [31], to researchgate.com and making its figure 2 freely available. Connor Shu is thanked for several large images that he stitched with Microsoft's Image Composite Editor together from the few translation periodic repeats in the original images in [34] so that plane symmetry group classifications with the author's information theory based approach could proceed in Fourier space.

## References

[1] M. I. Aroyo (Ed.), *International Tables for Crystallography, Volume A, Space-Group Symmetry*, 6th rev. ed., Chichester, UK: Wiley, 2016.

[2] H. Wondratschek and U. Müller (Eds.), *International Tables for Crystallography*, *Volume A1, Symmetry relations between space groups*, 1st ed., Chichester, UK: Kluwer, 2004.

[3] Th. Hahn (Ed.), *International Tables for Crystallography, Brief teaching edition of Volume A, Space-group Symmetry*, 5th rev. ed., Chichester, UK: Wiley & Sons, 2010.

[4] V. Kopský and D. B. Litvin (Eds.), *International Tables for Crystallography, Vol. E, Subperiodic groups*, 2nd ed., Chichester, UK: Wiley & Sons, 2010.

[5] E. Fedorov, "Симметрія на плоскости (La symétrie sur un plan)," *Proceedings of the Imperial Petersburg Mineralogical Society*, vol. 28, pp. 345–390, 1891, in Russian.

[6] G. Pólya, "Űber die Analogie der Kristallsymmetrie in der Ebene," *Zeitschrift für Kristallographie*, vol. 60, 278–282, 1924, in German.

[7] P. Moeck, "Crystallographic image processing for scanning probe and transmission electron microscopy," *Proc. 11th IEEE Intern. Conf. Nanotechnology*, Portland, OR, August 15-18, 2011, pp. 520–525, DOI: 10.1109/NANO.2011.6144304.

[8] P. Moeck, T. T. Bilyeu, J. Straton, M. Toader, M. Hietschold, U. Mazur, K. W. Hipps, and J. Rabe, "Crystallographic STM image processing for 2D periodic and highly symmetric molecule arrays," *Proc. 11th IEEE Intern. Conf. Nanotechnology*, Portland, OR, August 15-18, 2011, pp. 891–896, DOI: 10.1109/NANO.2011.6144508.

[9] P. Moeck, J. Straton, M. Toader, and M. Hietschold, "Crystallographic processing of scanning tunneling microscopy images of cobalt phthalocyanines on silver and graphite," *Mater. Res. Soc. Symp. Proc*. vol. 1318, 2011, pp. 148–154, DOI: 10.1557/opl.2011.278.

[10] P. Moeck, M. Toader, M. Abdel-Hafiez, and M. Hietschold, "Quantifying and enforcing two-dimensional symmetries in scanning probe microscopy images," in *Frontiers of Characterization and Metrology for Nanoelectronics*, D. G. Seiler, A. C. Diebold, R. McDonald, C. M. Garner, D. Herr, R. P. Khosla, and E. M. Secula, Eds., New York, NY, USA: American Institute of Physics, 2009, pp. 294–298, DOI: 10.1063/1.3251237.

[11] D. G. Morgan, Q. M. Ramasse, and N. D. Browning, "Application of two-dimensional crystallography and image processing to atomic resolution Z-contrast images," *J. Electron Microscopy*, vol. 58, pp. 223–244, 2009, DOI: 10.1093/jmicro/dfp007.

[12] N. D. Browning, J. P. Buban, M. Chi, B. Gipson, M. Herrera, D. J. Masiel, S. Mehraeen, D. G. Morgan, N. L. Okamoto, Q. M. Ramasse, B. W. Reed, and H. Stahlberg, "The application of scanning transmission electron microscopy (STEM) to the study of nanoscale systems," in *Modeling Nanoscale Imaging in Electron Microscopy*, T. Vogt, W. Dahmen, and P. Binev (Eds.), New York, NY, USA: Springer, 2012,(Nanostructure Science and Technology, Series Ed. D. J. Lockwood), DOI: 10.1007/978-1-4614-2191-7_2.

[13] X. Zou, S. Hovmöller, and P. Oleynikov, *Electron Crystallography: Electron Microscopy and Electron Diffraction*, IUCr Texts on Crystallography 16, Oxford, UK: Oxford University Press, 2011.

[14] X. Zou and S. Hovmöller, "Electron crystallography: imaging and single-crystal diffraction from powders," *Acta Cryst. A*, vol. 64, pp. 149–160, 2008, DOI: 10.1107/S0108767307060084, open access.

[15] P. Moeck, "Crystallographic image processing for scanning probe microscopy," in *Microscopy: Science Technology, Applications and Education*, Microscopy Book Series No. 4, Vol. 3, A. Méndez-Vilas and J. Diaz, Eds., Badajoz, Spain: Formatex Research Center, 2010, pp. 1951–1962, arXiv: 2012.08057.

[16] P. Moeck, A. Dempsey, and C. Shu, "Information theory based crystallographic symmetry classifications of a noisy 2D periodic scanning tunneling microscope image," *Microscopy and Microanalysis*, vol. 25 (Suppl. 2), pp. 184–185, 2019, (Proc. Annual Joint Meeting of the Microscopy Society and the Microanalysis Society of America, Portland, OR, August 4-8, 2019), DOI: 10.1017/S143192761900165X.

[17] P. Moeck, Towards more reasonable identifications of the symmetries in noisy digital images from periodic and aperiodic crystals, Proc. 21st International Conference on Nanotechnology, July 29 - 31, 2021, virtual, submitted.

[18] P. Moeck, "Advances in crystallographic image processing for scanning probe microscopy," in *Microscopy and Imaging Science: Practical approaches to applied research and education*, Microscopy Book Series No. 7, Méndez-Vilas, A. (Ed.), pp. 503–514, Badajoz, Spain: Formatex Research Center, 2017, arXiv: 2011.13102.

[19] J. C. Straton, B. Moon, T. T. Bilyeu, and P. Moeck, "Removal of multiple-tips artifacts from scanning tunneling microscope images by crystallographic averaging," *Adv. Struct. Chem. Imaging*, vol. 1, Art. no. 14 (12 pages), 2015, DOI: 10.1186/s40679-015-0014-6.

[20] J. C. Straton, T. T. Bilyeu, B. Moon, and P. Moeck, "Double-tip effects on scanning tunneling microscopy imaging of 2D periodic objects: Unambiguous detection and limits of their removal by crystallographic averaging in the spatial frequency domain," *Crystal Research and Technology*, vol. 49, pp. 663–680, 2014, DOI: 10.1002/crat.201300240

[21] P. Moeck, "Information theory approach to crystallographic symmetry classifications of noisy 2D periodic images," *Proc. 13th IEEE Conference on Nanotechnology Materials and Devices,* Portland, OR, 2018 (4 pages), DOI: 10.1109/NMDC.2018.8605865.

[22] P. Moeck, "Towards generalized noise-level dependent crystallographic symmetry classifications of more or less periodic crystal patterns," *Symmetry*, vol. 10, 2018, Art. no. 133 (46 pages), DOI: 10.3390/sym10050133, earlier version (before peer-review) at http://arxiv.org/abs/1801.01202, both in open access.

[23] P. Moeck and A. Dempsey, "Crystallographic symmetry classifications of noisy 2D periodic images in the presence of pseudo-symmetries of the Fedorov type," *Microscopy and Microanalysis*, vol. 25 (Suppl. 2), pp. 1936–1937, 2019, (Proc. Annual Joint Meeting of the Microscopy Society and the Microanalysis Society of America, Portland, OR, August 4-8, 2019), DOI: 10.1017/S1431927619010419.

[24] M. M. Julian, *Foundations of Crystallography with Computer Applications*, 2nd ed., Boca Raton, FL, USA: CRC Press, 2015.

[25] D. Schattschneider, "The plane symmetry groups: Their recognition and notation," *American Mathematical Monthly*, vol. 85, pp. 439–450, 1978, https://www.jstor.org/stable/2320063?origin=crossref&seq=1#metadata_info_tab_contents.

[26] Y. Liu, R. T. Collins, and Y. Tsin, "A computational model for periodic pattern perception based on frieze and wallpaper groups," *IEEE Transactions on Pattern Analysis and Machine Intelligence*, vol. 26, pp. 354–371, 2004, DOI: 10.1109/TPAMI.2004.1262332.

[27] M. Agustí-Melchor, Á. Rodas-Jordá, and J. M. Valiente-González, "Computational framework for symmetry classification of repetitive patterns," in *Communications in Computer and Information Science*, vol. 274, G. Csurka, M. Kraus, L. Mestetskiy, P. Richard, and J. Braz, (Eds.), Berlin, Heidelberg, Germany: Springer-Verlag, 2013, pp. 257–270, DOI: 10.1007/976-3-642-32350-8_18, (Proc. Intern. Conference on Computer Vision, Imaging and Computer Graphics, VISIGRAPP 2011, CCIS 274).

[28] M. Agustí-Melchor, A. Rodas-Jordá, and J. M. Valiente-González, "Classification of Repetitive Patterns Using Symmetry Group Prototypes," in *Pattern Recognition and Image Analysis*, J. Vitrià, J. M. Sanches, and M. Hernández (Eds.), Berlin, Heidelberg, Germany: Springer, 2011, Lecture Notes in Computer Science, vol. 6669, pp. 84–91, DOI: 10.1007/978-3-642-21257-4_11, (Proc. IbPRIA 2011).

[29] M. Agustí-Melchor, A. Rodas-Jordá, and J. M. Valiente-González, "Computational symmetry via prototype distances for symmetry groups classification," *Proc. Intern. Conf. Computer Vision Theory and Applications*, 2011, pp. 85–94, DOI: 10.5220/0003375300850093

[30] M. Valor, F. Albert, J. M. Gomis, and M. Contero, "Analysis tool for cataloguing textile and tile pattern designs," *Proc. 2003 Intern. Conf. Computational Science and its Applications*, V. Kumar, M. L. Gavrilova, C. J. K. Tan, and P. L'Ecuyer, (Eds.), Berlin, Heidelberg, Germany: Springer, 2003, Lecture Notes in Computer Science, vol. 2669, Part III, pp. 569–578, DOI: 10.1007/3-540-44842-X_58, (Proc. ICCSA'03).

[31] M. Valor, F. Albert, J. M. Gomis, and M. Contero, "Textile and tile pattern design automatic cataloguing using detection of the plane symmetry group," *Proc. Computer Graphics International*, pp. 112–119, 9-11 July 2003, DOI: 10.1109/CGI.2003.1214455.

[32] J. M. Valiente, F. Albert, and J. M. Gomis, "A computational model for pattern and tile designs classification using plane symmetry groups," in *Progress in Pattern Recognition, Image Analysis and Applications*, M. Lazo, A. Sanfeliu, and M. L. Cortés (Eds.), Berlin, Heidelberg, Germany: Springer, 2005, Lecture Notes in Computer Science, vol. 3773, pp. 849–860, DOI: 10.1007/11578079_88, (Proc. CIARP 2005).

[33] J. M. Valiente González, F. E. Albert Gil, and J. M. Gomis Martí, "Feature extraction and classification of textile images: Towards a design information system for the textile industry," *Proc. 2nd International Workshop on Pattern Recognition in Information Systems*, J. M. Iñesta Quereda and M. L. Micó Andrés (Eds.), pp. 77–94, (Proc. PRIS 2002, Alicante, Ciudad Real, April 2002).

[34] F. Albert, J. M. Gómis, J. Blasco, J. M. Valiente, and N. Aleixos, "A new method to analyse mosaics based on symmetry group theory applied to Islamic geometric patterns," *Computer Vision and Image Understanding*, vol. 130, pp. 54–70, 2015, DOI: 10.1016/j.cviu.2014.09.002.

[35] Y. Liu, H. Hel-Or, C. S. Kaplan, and L. Van Gool, "Computational symmetry in computer vision and computer graphics", *Foundations and Trends® in Computer Graphics and Vision*, vol. 5, pp. 1–195, 2009, DOI: 10.1561/0600000008.

[36] E. Makovicky and M. Ghari, "Neither simple nor perfect: from defect symmetries to conscious pattern variations in Islamic ornamental art," *Symmetry: Culture and Science*, vol. 29, pp. 279–301, 2018, DOI: 10.26830/symmetry_2018_2_279.

[37] E. Makovicky, "Comment on 'decagonal and quasi-crystalline tilings in medieval Islamic architecture'," *Science*, vol. 318, Art. no. 1383a (2 pages), 2007, DOI: 10.1126/science.1146262.

[38] K. Kanatani, "Geometric information criterion for model selection," *International Journal of Computer Vision*, vol. 26, pp. 171–189, 1998, DOI: 10.1023/A:1007948927139.

[39] K. Kanatani, "Uncertaintly modeling and model selection for geometric inference," *IEEE Transactions on Pattern Analysis and Machine Intelligence*, vol. 26, pp. 1307–1319, 2004, DOI: 10.1109/TPAMI.2004.93.

[40] K. Kanatani, "Uncertainty modeling and geometric inference," in *Handbook of Geometric Computing: Applications in Pattern Recognition, Computer Vision, Neuralcomputing, and Robotics*; E. B. Corrochano, Ed., Berlin, Germany: Springer: 2005, Chapter 14, pp. 462–491.

[41] K. Kanatani, *Statistical Optimization for Geometric Computation: Theory and Practice*, Mineola, NY, USA: Dover Publ., 2005.

[42] K. Kanatani, "Comments on 'Symmetry as a Continuous Feature'," *IEEE Transactions on Pattern Analysis and Machine Intelligence*, vol. 19, pp. 246–247, 1997, DOI: 10.1109/34.584101.

[43] E. V. Chuprunov, "Fedorov pseudosymmetry of crystals: Review," *Crystallography Reports*, vol. 52, pp. 1–11, 2007, DOI: 10.1134/S1063774507010014, (original in Russian: *Kristallografiya*, vol. 52, No. 1, pp. 5–16, 2007).

[44] P. Moeck and P. DeStefano, "Accurate lattice parameters from 2D periodic images for subsequent Bravais lattice type assignments," *Adv. Struct. Chem. Imaging*, vol. 4, Art. no. 5 (33 pages), 2018, DOI: 10.1186/s40679-018-0051-z, open access.

[45] R. Zabrodsky, S. Peleg, and D. Avnir, "Symmetry as a continuous feature," *IEEE Transactions on Pattern Analysis and Machine Intelligence*, vol. 17, pp. 1154–1166, 1995, DOI: 10.1109/34.476508.

[46] R. K. Vasudevan, N. Laanait, E. M. Ferragut, K. Wang, D. B. Geohegan, K. Xiao, M. Ziatdinov, S. Jesse, O. Dyck, and S. V. Kalinin, "Mapping mesoscopic phase evolution during E-beam induced transformations via deep learning of atomically resolved images," *npj Computational Materials*, vol. 30 (9 pages and on-line supporting material), 2018, DOI: 10.1038/s41524-018-0086-7.

[47] W. B. Park, J. Chung, J. Jung, K. Sohn, S. P. Singh, M. Pyo, N. Shin, and K.-S. Sohn, "Classification of crystal structure using a convolutional neural network," *IUCrJ*, vol. 4, pp. 486–494, 2017, DOI: 10.1107/S205225251700714X, open access.

[48] A. Ziletti, D. Kumar, M. Scheffler, and L. M. Ghiringhelli, "Insightful classifications of crystal structures using deep learning," *Nature Communications*, vol. 9, Art. no 2775 (9 pages), 2018, DOI: 10.1038/s41467-018-05169-6, open access.

[49] P. Mungofa, A. Schumann, and L. Waldo, "Chemical crystal identification with deep learning machine vision," *BMC Research Notes*, 11, Art. no 703 (6 pages), 2018, DOI: 10.1186/s13104-018-3813-8.

[50] R. K. Vasudevan, private communication, 2018.

[51] A. Belsky, M. Hellenbrandt, V. L. Karen, and P. Luksch, "New developments in the Inorganic Crystal Structure Database (ICSD): accessibility in support of materials research and design," *Acta Cryst. B*, vol. 58, pp. 364–369, 2002, DOI: 10.1107/S0108768102006948.

[52] S. Curtarolo, W. Setyawan, S. Wang, J. Xue, K. Yang, R. H. Taylor, L. J. Nelson, G. L. W. Hart, S. Sanvito, M. Buongiorno-Nardelli, N. Mingo, and O. Levy, "AFLOWLIB.ORG: A distributed materials properties repository from high-throughput *ab initio* calculations," *Computational Materials Science*, vol. 58, pp. 227–235, 2012, DOI: 10.1016/j.commatsci.2012.02.002.

[53] R. W. Grosse-Kunstleve, "Algorithms for deriving crystallographic space group information," *Acta Cryst. A*, vol. 55, pp. 383–395, 1999, DOI: 10.1107/S0108767301016658.

[54] M. De Graef, and M. E. McHenry, *Structure of Materials: An Introduction to Crystallography, Diffraction, and Symmetry*, 2nd rev. ed., Cambridge, UK: Cambridge University Press, 2012.

[55] P. Terpstra and L. W. Codd, *Crystallometry*, New York, NY, USA: Academic Pres, 1961.

[56] A. K. Boldyrew and W. W. Doliwo-Dobrowolsky, *Bestimmungstabellen für Kristalle (Определитель Кристаллов)*. vol. I, part 1, Einleitung, Tetragyrische Syngonie, Zentrales Wissenschaftliches Institut der Geologie und Schürfung, Leningrad and Moscow, 1937 (in Russian and German).

[57] W. W. Doliwo-Dobrowolsky and G. P. Preobraschensky, *Bestimmungs-tabellen für Kristalle (Определитель Кристаллов)*. vol. I, part 2, Trigyrische and Hexagyrische Syngonien allgemeine Ergänzungen zu den mittleren Syngonien, Zentrales Wissenschaftliches Institut der Geologie und Schürfung, Leningrad and Moscow, 1939 (in Russian and German).

[58] S. Lemonick, "Is machine learning overhyped?," *Chemical and Engineering News*, 2018, https://cen.acs.org/physical-chemistry/computational-chemistry/machine-learning-overhyped/96/i34.

[59] P. Moeck, A. Dempsey, and C. Shu, so far unpublished.

[60] Microsoft ICE 2.0.3, Image Composite Editor, for Windows 7, 8, and 10, freely downloadable at https://www.microsoft.com/en-us/research/product/computational-photography-applications/image-composite-editor/.

[61] S. Park and C. F. Quate, "Digital filtering of scanning tunneling microscope images," *J. Appl. Phys.*, vol. 62, pp. 312–314, 1987, DOI: 10.1063/1.339150.

[62] Y. Liu, http://vivid.cse.psu.edu/.

[63] B. Wichmann, http://www.tilingsearch.org/.

[64] F. A. Farris, *Creating Symmetry: The artful mathematics of wallpaper patterns*, Princeton, NJ, USA: Princeton University Press, 2015.

[65] E. Malkovicky, *Symmetry through the eyes of the old masters*, Berlin: De Gruyter, 2016.

[66] D. Schattschneider, *Visions of symmetry: notebooks, periodic drawings, and related work of M. C. Escher*, New York, NY, USA: Freeman, 1990.

[67] P. J. Morandi, *The Classification of Wallpaper Patterns: From Group Cohomology to Escher's Tessellations*, Department of Mathematical Sciences, New Mexico State University, Las Cruces, NM 88003, free on-line book, http://sierra.nmsu.edu/morandi/notes/wallpaper.pdf and https://drive.google.com/file/d/12fti6FKkwqlCyntRGAkXtjbrfUyw3p-g/view, accessed February 4, 2019.

[68] K. Kanatani, "Geometric BIC," *Memories of the Faculty of Engineering, Okayama University*, vol. 42, pp. 10–17, January 2008.

[69] K. P. Burnham and D. R. Anderson, *Model Selection and Multimodel Inference: A Practical Information-Theoretic Approach*, New York, NY, USA: Springer, 2002.

[70] D. R. Anderson, *Model based inference in the life sciences: A primer on evidence*, New York, NY, USA: Springer, 2008.

[71] M. Nespolo, M. I. Aroyo, and B. Souvignier, "Crystallographic shelves: space-group hierarchy explained," *J. Appl. Cryst.*, vol. 51, pp. 1481–1491, 2018, DOI: 10.1107/S1600576718012724.

[72] P. Moeck, A. Dempsey, and C. Shu, so far unpublished.

[73] P. Moeck, A. Dempsey, and C. Shu, so far unpublished.

[74] S. Hembacher, F. J. Giessibl, J. Mannhart, and C. F. Quate, "Revealing the hidden atom in graphite by low-temperature atomic force microscopy," *PNAS*, vol. 100, pp. 12539–12542, 2003, DOI: 10.1073/pnas.2134173100.

[75] D. Tománek and S. G. Lourie, "First-principle calculations of highly asymmetric structure in scanning-tunneling-microscopy images of graphite," *Phys. Rev. B*, vol. 37, pp. 8327–8336, 1988, DOI: 10.1103/PhysRevB.37.8327.

[76] H. Akaike, “A new look at the statistical model identification,” *IEEE Trans. Automatic Control.*, vol. 19, pp. 716–723, 1974, DOI: 10.1109/TAC.1974.1100705.

[77] T. C. Chamberlin, “The method of multiple working hypotheses”, *Science*, vol. 15, pp. 92–96, 1890, reprinted *Science*, vol. 754–759, 1965, DOI: 10.1126/science.148.3671.754, https://www.jstor.org/stable/1716334.